\documentclass[%
aip,
 amsmath,amssymb,
 reprint,%
nofootinbib]{revtex4-1}

\usepackage{graphicx}
\usepackage{dcolumn}
\usepackage{bm}
\usepackage[usenames]{color}
\usepackage[dvipsnames]{xcolor}
\usepackage[utf8]{inputenc}
\usepackage[T1]{fontenc}
\usepackage{mathptmx}
\usepackage{soul}
\begin{document}

\preprint{AIP/123-QED}

\title{Influence of astrocyte feedback and noise on breathing neural clusters}

\author{Ediline L.F. Nguessap}%
\email{fonela@usp.br}
\affiliation{Department of Physics-FFCLRP, University of S\~ao Paulo, Ribeirao Preto-SP, 14040-901, Brasil.}

\author{Thierry Njougouo}%
\affiliation{IMT School for Advanced Studies, Piazza San Francesco 19, 55100 Lucca, Italy}
\affiliation{Department of Physics-FFCLRP, University of S\~ao Paulo, Ribeirao Preto-SP, 14040-901, Brasil.}

%

\author{Fernando F. Ferreira}%
\affiliation{Department of Physics-FFCLRP, University of S\~ao Paulo, Ribeirao Preto-SP, 14040-901, Brasil.}

\date{\today}

\begin{abstract}
Astrocytes are increasingly recognized as active regulators of neural dynamics, yet their role in large-scale synchronization remains incompletely understood. Building on the discovery of breathing cluster synchronization in neuron–astrocyte networks, we investigate whether this phenomenon persists under more biologically realistic conditions by incorporating thermosensitive neuronal dynamics, intrinsic electric-field interactions, stochastic perturbations, and delayed astrocytic feedback. Through systematic numerical simulations of a network containing structurally symmetric neuronal subsets, we show that breathing clusters characterized by intermittent alternations between synchronized and desynchronized states, remain robust across a broad range of biophysical conditions. The results reveal a non-monotonic dependence on astrocytic attenuation and feedback parameters, with intermediate values maximizing rhythmic switching through a slow push–pull regulatory mechanism. Temperature and electric-field effects primarily modulate neuronal excitability and the tolerance of the breathing state to noise, while elevated stochastic perturbations progressively destabilize cluster coherence. In contrast, the phenomenon exhibits remarkable robustness to biologically realistic astrocytic delays. Importantly, our findings indicate that the emergence of breathing synchronization remains fundamentally governed by the interplay between network symmetry and astrocyte-mediated feedback, whereas temperature, electric fields, and noise act as modulators of its stability and temporal organization. These results establish breathing cluster synchronization as a robust organizing principle of astrocyte-regulated networks and provide a quantitative framework for understanding how glial regulation shapes cortical state-switching under realistic physiological conditions.
\end{abstract}

\maketitle

\begin{quotation}
Understanding how interactions between neurons and astrocytes shape collective synchronization dynamics is a fundamental challenge in nonlinear neuroscience. Here, we demonstrate that breathing cluster synchronization, characterized by rhythmic alternation between synchronized and desynchronized neuronal groups, persists under thermosensitive dynamics, stochastic perturbations, electric-field interactions, and delayed astrocytic regulation. Through systematic numerical simulations, we show that the phenomenon is primarily controlled by the interplay between network symmetry and adaptive astrocyte-mediated feedback, while environmental factors mainly influence its robustness and temporal structure. These results identify breathing clusters as a robust emergent state of adaptive neural networks and demonstrate that astrocyte-mediated feedback plays a central role in regulating collective synchronization dynamics.
\end{quotation}

\section{Introduction}

Coordinated neural activity constitutes a cornerstone of information processing within the brain, supporting essential cognitive operations such as sensory integration, memory formation, and attentional focus\cite{singer1999neuronal,varela2001brainweb,ward2003synchronous,uhlhaas2006neural,hipp2011oscillatory,ma2015review,wu2022understanding}. While traditional computational neuroscience has prioritized neuronal mechanisms to explain these rhythms, the brain is composed of a heterogeneous cellular environment where glial cells significantly outnumber neurons\cite{barres2008mystery,kettenmann2008neuroglia}.  Consequently, understanding collective neural behavior requires moving beyond neuron-centric models to incorporate the regulatory influence of the glial network, particularly within the context of complex network topology\cite{watts1998collective,barabasi1999emergence,sporns2004small,eguiluz2005scale,bassett2017network}.

For decades, the neurocentric view dominated the field, positing that glial cells served merely as passive structural scaffolding or metabolic support for active neuronal signaling\cite{kandel2000principles,patro2022biology}. Contemporary research, however, has precipitated a paradigm shift, establishing astrocytes as active participants in information processing and synaptic regulation\cite{haydon2001glia,fields2002new}. As the most prevalent glial subtype, astrocytes envelop synapses to form the tripartite synapse, where they actively modulate neurotransmitter concentration and synaptic efficacy\cite{cornell1990glutamate,parpura1994glutamate,parri2001spontaneous,perea2007astrocytes,ota2013role,kastanenka2020roadmap}. Crucially, in vivo evidence indicates that astrocytic gliotransmission, specifically the uptake and release of glutamate, can drive transient synchronization events and regulate the switching between cortical UP and DOWN states\cite{fellin2004neuronal,angulo2004glutamate,poskanzer2011astrocytic,poskanzer2016astrocytes}. Despite these experimental advances, a comprehensive theoretical framework linking astrocytic feedback to large-scale network synchronization remains incomplete\cite{tang2017astrocyte}.

Existing computational models of neuron-astrocyte interactions have largely been confined to minimal motifs, such as single synapses or small clusters involving one astrocyte and a few neurons\cite{nadkarni2003spontaneous,allegrini2009influence,pereira2009role,tang2013information,makovkin2020astrocyte,chen2024astrocyte,chen2025neural}. While these reduced models successfully demonstrate local modulatory effects, they fail to capture the emergent dynamics of  cortical networks where topology plays a critical role\cite{amiri2011role,amiri2013astrocyte,araque1999tripartite,amiri2012functional,postnov2009dynamical,gordleeva2021modeling}. A significant step forward was made by Wang et al.\cite{wang2023breathing}, who introduced a complex network model demonstrating breathing cluster synchronization defined as a state where symmetric neuronal subsets intermittently switch between coherence and incoherence due to astrocytic feedback. However, this foundational work relied on deterministic dynamics and simplified neuronal models, omitting critical biophysical variables present in biological tissue\cite{steriade2001natural,cossart2003attractor,buzsaki2004neuronal}.

Real neural systems operate within a fluctuating physical environment where intrinsic and extrinsic factors continuously perturb dynamics. For instance, neuronal excitability is inherently sensitive to temperature variations, which alter ion channel kinetics and recovery processes\cite{xu2020dynamics,hussain2021chimera,zhu2021effects}. Similarly, endogenous and exogenous electric fields interact with neuronal membranes via ephaptic coupling, offering a non-synaptic pathway for synchronization control\cite{bikson2004effects,radman2007spike,simo2021chimera}. Furthermore, biological networks are subject to stochastic noise arising from channel fluctuations and synaptic variability\cite{matsumoto1983noise,zhou2002noise}. The interplay between these environmental factors and slow astrocytic feedback loops remains unexplored in the context of cluster synchronization\cite{ghosh2022synchronized,de2008adaptive}.

In this work, we extend the framework of neuron-astrocyte networks by integrating thermosensitive neuronal dynamics, intrinsic electric field variables, and stochastic perturbations. We investigate how these biophysical factors influence the stability and temporal structure of breathing clusters. Our analysis reveals that while astrocytic feedback drives the intermittent synchronization mechanism, the rhythm and robustness of these clusters are finely tuned by temperature, field strength, and noise intensity. These findings provide a more biologically realistic description of state-switching in the neocortex and suggest that astrocytes function as adaptive controllers capable of maintaining coordination amidst environmental variability\cite{nguessap2026modulation,nguessap2026dynamic,rakshit2019transitions}.

\section{Model}
\label{sec:model}

The proposed model integrates three key components: (1) a thermosensitive neuron model augmented with an intrinsic electric field variable, (2) chemical synaptic coupling within a random network architecture, and (3) astrocytic modulation of synaptic strength based on network-level activity.

The neuronal membrane, treated as a charged surface with uniform charge density $\sigma = q/S$, generates an electric field $E = \sigma/(2\varepsilon_1)$, where $\varepsilon_1$ is the dielectric constant of the surrounding medium and $S$ represents the surface area of the neuronal membrane. The corresponding membrane potential difference is $\Delta V \approx E\sqrt{S}$. In the nondimensional formulation used here, this physical membrane potential is represented by the dimensionless variable \(x\), with the geometric factor $\sqrt{S}$ which is proportional to the radius of the cell (considering as a ball shape) and the dielectric constant are absorbed into the effective coupling parameters (such as $r$ and $k$). This relation provides the biophysical motivation for dynamically coupling the electric field variable $E$ to the recovery variable $y$ (for a detailed derivation, see Xu et al.~\cite{xu2020dynamics} and Nguessap et al.~\cite{nguessap2026modulation}), rather than appearing as an explicit algebraic term.
The electric-field variable is introduced to capture ephaptic-like modulation of neuronal excitability. In the present work, it is used primarily to assess whether breathing synchronization remains robust under additional biophysical interactions rather than as the primary mechanism generating breathing dynamics.
The governing equations are:
\begin{equation}
\begin{cases}
\displaystyle\frac{dx}{dt} = x(1 - \xi) - \frac{1}{3}x^3 - y + I + A\cos(\omega t), \\[8pt]
\displaystyle\frac{dy}{dt} = c\left[x + a - b e^{1/T}y\right] + rE, \\[8pt]
\displaystyle\frac{dE}{dt} = ky,
\end{cases}
\label{eq:single_neuron}
\end{equation}
where $x$ represents the membrane potential, $y$ is a recovery variable encapsulating slow ion currents (primarily potassium), and $E$ captures the intrinsic electric field generated by transmembrane ion movements. The temperature dependence enters through the term $b e^{1/T}$, with higher temperatures increasing the rate of recovery current decay. The parameter $r$ scales the coupling between the electric field and ionic currents, while $k$ determines how currents generate field variations. External periodic stimulation is modeled via $A\cos(\omega t)$, and a constant bias current $I$ sets the baseline excitability.

To model external field influences, such as those from transcranial stimulation or endogenous field effects, we extend the field dynamics to include an applied component $E_{ext}$:
\begin{equation}
\frac{dE}{dt} = ky + E_{ext}, 
\label{eq:external_field}
\end{equation}
where $E_{ext}=D\eta(t)$ represents an external electric field, modeled as a  noise perturbation  of amplitude $D$ with $\eta $ a Gaussian white noise.
\begin{figure}
    \centering  \includegraphics[width=0.6\linewidth]{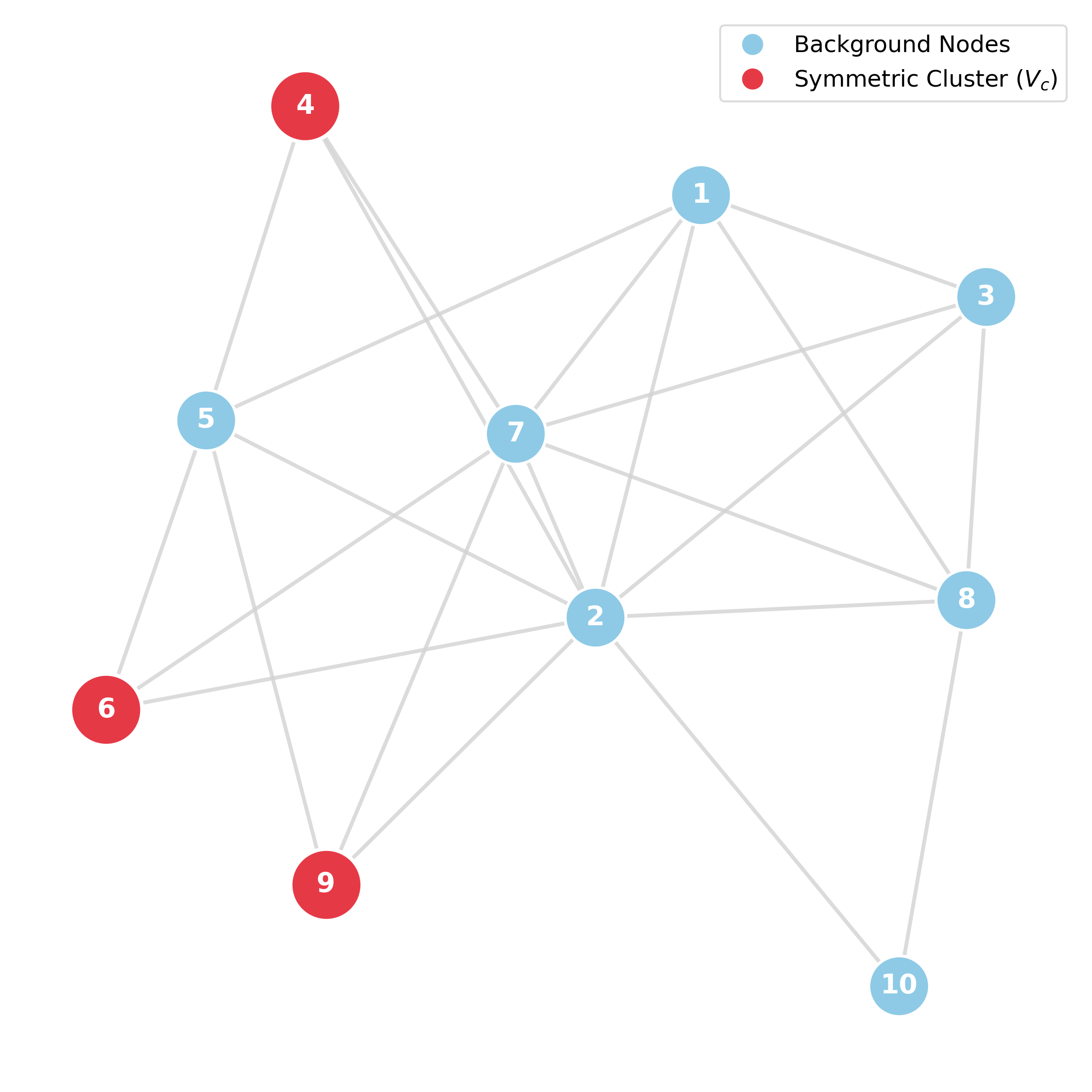}
    \caption{Network topology with symmetric cluster $V_c = \{4,6,9\}$ (red nodes) embedded in a random background network (blue nodes). Gray edges denote bidirectional synaptic connections. }
    \label{net}
\end{figure}
We consider a network of $N$ neurons arranged in a random topology that explicitly incorporates a structurally symmetric cluster, which is a prerequisite for the emergence of breathing cluster synchronisation (Fig.~\ref{net}) . Following the principle of structural equivalence, a subset of nodes $V_c \subset \{1, \dots, N\}$ of size $m$ is defined as symmetric if they share identical incoming connection patterns from the rest of the network. Mathematically, for any $i, j \in V_c$, the adjacency matrix elements satisfy {$G_{ki} = G_{kj}$ for all $k \notin \{i, j\}$.

The network is constructed by first generating a random background connectivity for all node pairs with probability of $ 0.7$. For the $m$ symmetric nodes in $V_c$, we assign an identical binary input vector from the remaining $N-m$ nodes, while setting intra-cluster connections to zero. The network size is kept small ($N=10$) in order to isolate the dynamical mechanism underlying breathing cluster synchronization and to facilitate systematic exploration of the parameter space. The present model should therefore be regarded as a proof-of-principle framework rather than a quantitative representation of cortical-scale networks. The adjacency matrix is then symmetrized ($G = (G + G^*)/2$, with $G^*$ the transpose matrix) and the diagonal zeroed ($G_{ii} = 0$) to ensure bidirectional coupling without self-loops.

Crucially, the breathing phenomenon which is characterized by intermittent transitions between coherent and incoherent states--emerges specifically within this symmetric subset $V_c$. Because the global astrocytic feedback acts uniformly on these structurally equivalent nodes, they oscillate collectively. Consequently, the cluster synchronization error $\delta x_{clu}(t)$ and the desynchronization fraction $P$ (defined later in Eq.~\ref{eq:P_measure}) are computed exclusively over the symmetric cluster $V_c$ (e.g., $V_c = \{4, 6, 9\}$ for $m=3$) to accurately quantify this rhythmic alternation. The dynamics of neuron $i$ become:
\begin{equation}
\begin{cases}
\displaystyle\frac{dx_i}{dt} = x_i(1 - \xi) - \frac{1}{3}x_i^3 - y_i + I + A\cos(\omega t) + C_i, \\[8pt]
\displaystyle\frac{dy_i}{dt} = c\left[x_i + a - b e^{1/T}y_i\right] + rE_i, \\[8pt]
\displaystyle\frac{dE_i}{dt} = ky_i + E_{ext},
\end{cases}
\label{eq:network}
\end{equation}
where the synaptic input $C_i$ incorporates chemical coupling:
\begin{equation}
C_i = {\epsilon} (x_s - x_i) \sum_{j=1}^{N} G_{ij} \Gamma(x_j), \quad 
\Gamma(x_j) = \frac{1}{1 + e^{-\lambda(x_j - \theta_s)}}.
\label{eq:coupling}
\end{equation}
Here $G_{ij}$ is the adjacency matrix of the network $G_{ij}=1$ if neurons $i$ and $j$ are synaptically connected), $x_s=2$ mV is the synaptic reversal potential, $\theta_s = -0.25$ mV sets the firing threshold for synaptic release, and $\lambda=10$ controls the steepness of the sigmoidal activation function $\Gamma(\cdot)$. The key variable $\epsilon(t)$ represents the global synaptic strength, which is dynamically regulated by astrocytic activity.

Astrocytes sense neuronal activity via glutamate spillover and modulate synaptic efficacy in a feedback manner. To account for the activity-dependent nature of astrocyte-mediated
regulation, we allow the effective coupling strength $\epsilon$ to evolve
dynamically in response to the collective state of the neuronal network.
This activity-dependent feedback is modeled through a first-order
differential equation with a delayed astrocytic response: 
\begin{equation}
\dot{\epsilon} = -\alpha \epsilon + \beta R(t - \tau) + \gamma.
\label{eq:astro}
\end{equation}

The attenuation coefficient $\alpha$ determines the decay rate of astrocytic influence, $\beta$ quantifies the strength of the dynamical adaptive feedback coupling from neural synchrony to astrocytic output, $\gamma$ represents a constant basal glutamate release, and $\tau$ is a delay accounting for the slow calcium dynamics in astrocytes.  The order parameter $R(t)$ measures the degree of phase coherence in the network:
\begin{equation}
R(t) = \frac{1}{N} \left| \sum_{j=1}^{N} e^{i\theta_j(t)} \right|, \quad  \text{with} \quad i^2 = -1,
\label{eq:order}
\end{equation}
where the instantaneous phase $\theta_j(t)$ of neuron $j$ is linearly interpolated between successive spikes occurring at times $t_{j,k}$ and $t_{j,k+1}$:
\begin{equation}
\theta_j(t) = 2\pi \frac{t - t_{j,k}}{t_{j,k+1} - t_{j,k}}, \quad t_{j,k} \le t \le t_{j,k+1}.
\end{equation}
The dependence of the astrocytic drive on $R(t)$ reflects the biological observation that synchronized neuronal firing produces larger extracellular glutamate transients, which in turn elicit stronger astrocytic calcium responses and gliotransmitter release \cite{perea2007astrocytes,cresswell2018compartmental,rusakov2025active}.
To quantify synchronization, we compute the global synchronization error:
\begin{equation}
\delta x_{\text{net}}(t) = \frac{1}{N} \sum_{i=1}^{N} |x_i(t) - \bar{x}(t)|, \quad \text{with}\quad \bar{x}(t) = \frac{1}{N} \sum_{i=1}^{N} x_i(t),
\label{eq:global_error}
\end{equation}
and the cluster synchronization error for a subset $V_c = \{4,6,9\}$ of neurons sharing symmetric inputs:
\begin{equation}
\delta x_{\text{clu}}(t) = \frac{1}{|V_c|} \sum_{l \in V_c} |x_l(t) - \hat{x}(t)|, \quad \text{with}\quad \hat{x}(t) = \frac{1}{|V_c|} \sum_{l \in V_c} x_l(t).
\label{eq:cluster_error}
\end{equation}
The breathing cluster phenomenon is characterized by periodic oscillations in $\delta x_{\text{clu}}(t)$. We quantify the prevalence of desynchronization by measuring the fraction of time $P$ during which the cluster error exceeds a threshold $\delta_{\text{th}} = 0.01$:
\begin{equation}
P = \frac{1}{T_{sim}} \int_0^{T_{sim}} \Theta\big(\delta x_{\text{clu}}(t) - \delta_{\text{th}}\big) \, dt,
\label{eq:P_measure}
\end{equation}\
where $\Theta(\cdot)$ is the Heaviside step function. 
The desynchronization fraction $P$ therefore provides a complementary
quantitative measure of the temporal prevalence of desynchronization,
whereas the temporal evolution of $\delta x_{\mathrm{clu}}(t)$ is used to
distinguish rigid synchronization, breathing synchronization, and
persistent desynchronization.

Unless otherwise noted, simulations use the following fixed parameters: $a=0.7$, $c=0.1$, $\xi=0.175$, $T=5$, $b=0.4$, $I=0.5$, $k=0.001$, $\omega=0.005$, $r=0.001$; astrocyte parameters are typically set to $\alpha = 3\times10^{-2}$, $\beta = 8\times10^{-3}$, $\gamma = 2\times10^{-3}$, $\tau = 200$. 

\section{Results}
\label{sec:results}
Numerical simulations were performed using a fifth-order Runge--Kutta integration scheme with a fixed time step $\Delta t = 0.01$. Time is expressed in dimensionless units, and each simulation consists of $n = 5 \times 10^5$ integration steps, corresponding to a total simulation duration of $T_{\mathrm{sim}} = n\Delta t = 5000$. Delayed astrocytic feedback was implemented through a discrete buffer storing
past values of the global order parameter $R(t)$.
The initial values of the state variables $(x_i,y_i,E_i)$ were randomly drawn from a uniform distribution in $[-1,1]$, while the initial astrocyte-regulated synaptic strength $\varepsilon(0)$ was randomly selected from the interval
$[0,1]$. Additive Gaussian white noise was independently generated at each integration step to model stochastic perturbations.


\subsection{Emergence of Breathing Cluster Synchronization}
\begin{figure}
\centering
    \begin{tabular}{cc}
       (a) & (b)\\     \includegraphics[width=0.235\textwidth]{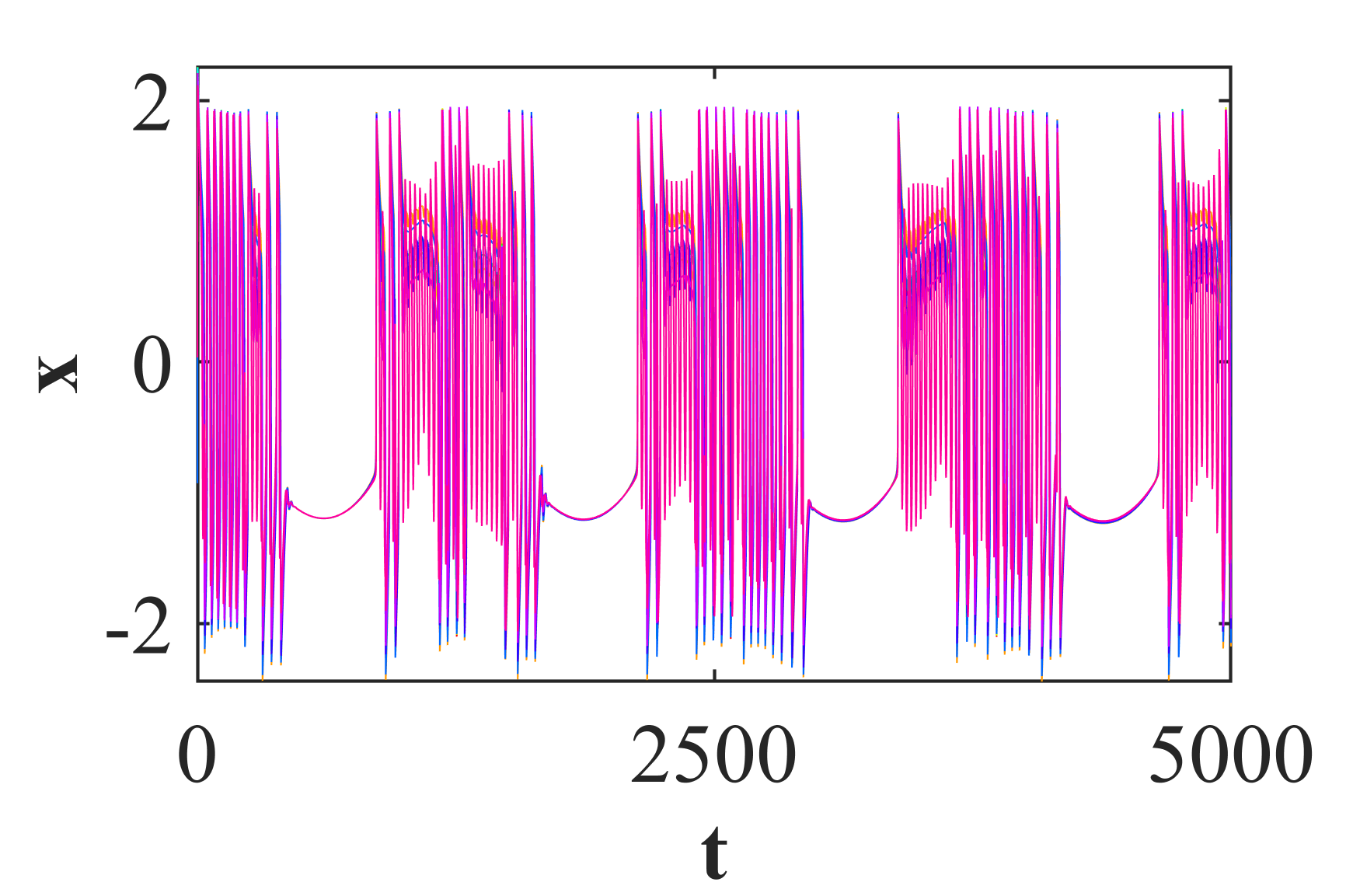} &         \includegraphics[width=0.235\textwidth]{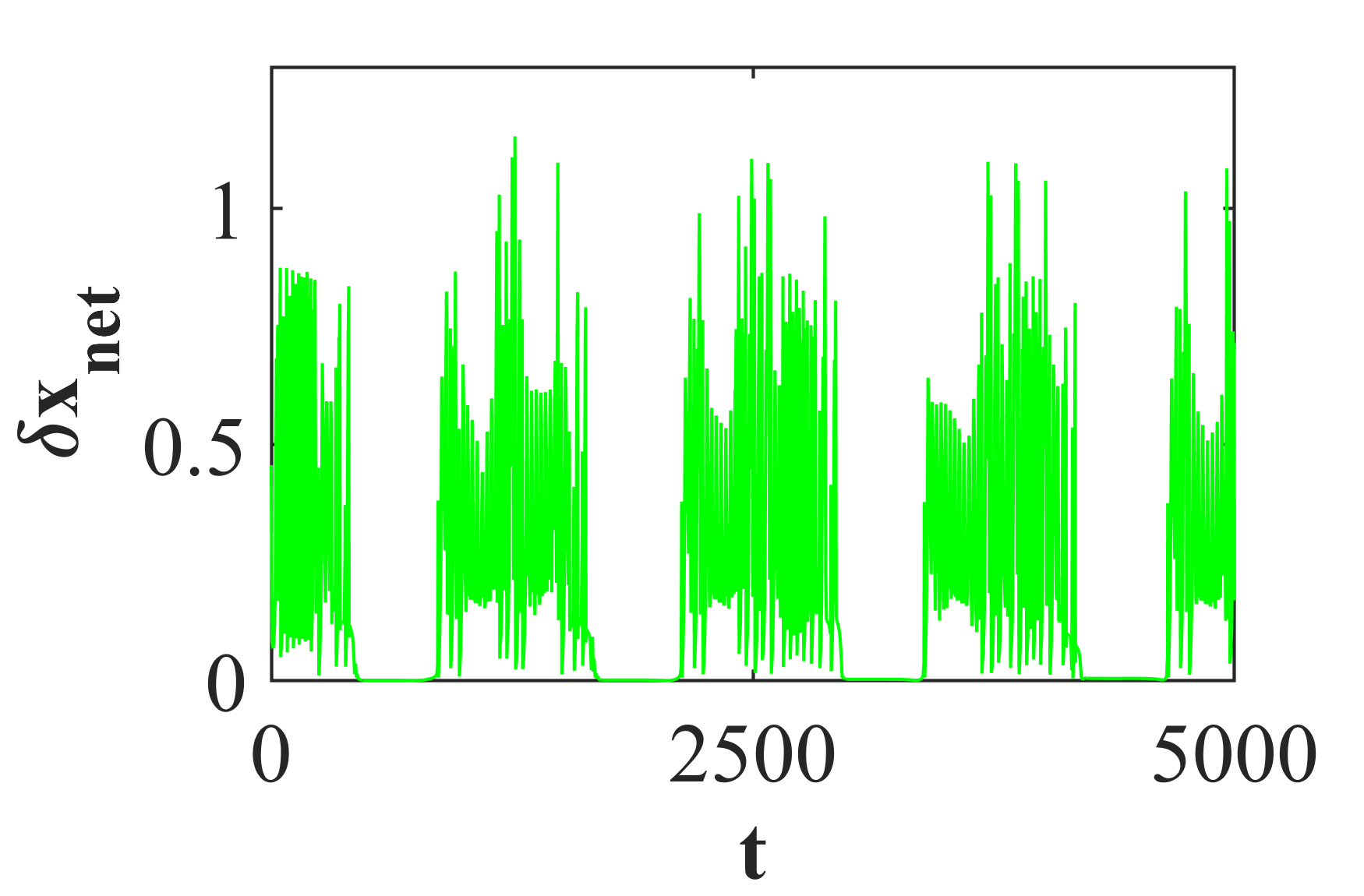}\\
       (c) & (d)\\       \includegraphics[width=0.235\textwidth]{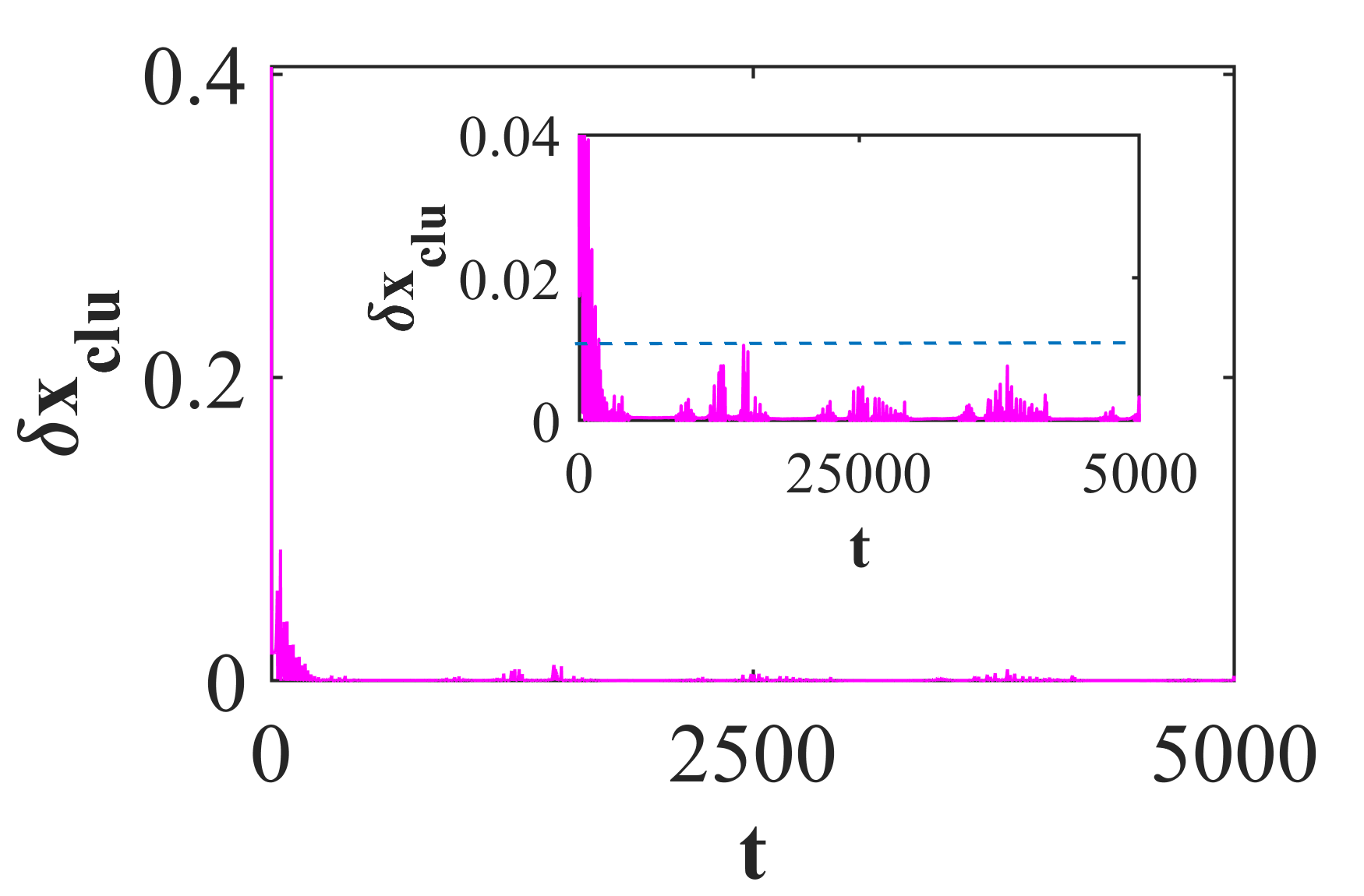}&
     \includegraphics[width=0.235\textwidth]{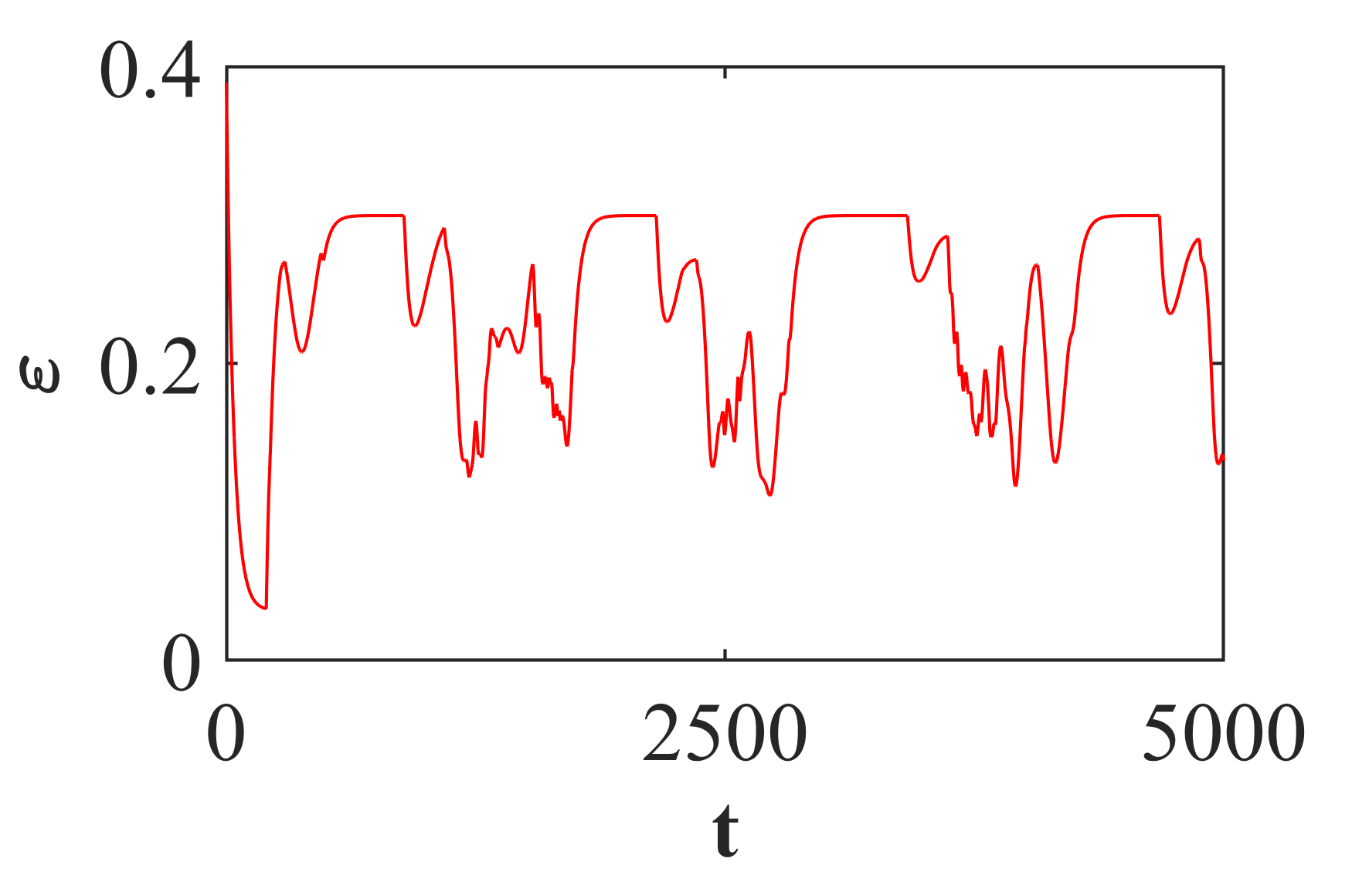}
    \end{tabular}
    \caption{Time evolution of the network under low noise (\(D = 0.01\)). (a) Membrane potentials \(x_i(t)\) of all neurons, showing clustered firing patterns. (b) Global synchronization error \(\delta x_{\text{net}}(t)\). (c) Cluster synchronization error \(\delta x_{\text{clu}}(t)\) for the symmetric neuron subset \(V_c = \{4,6,9\}\). The blue dashed line in the $\delta x_{\text{clu}}(t)$ panels indicates the desynchronization threshold $\delta_{\text{th}} = 0.01$. (d) Synaptic coupling strength \(\epsilon(t)\) modulated by astrocyte feedback. Parameters: \(\alpha = 0.03\), \(\beta = 0.008\), \(\gamma = 0.002\).}
    \label{fig:1}
\end{figure}
To illustrate how the adaptive astrocyte-mediated feedback shapes the collective dynamics of the neuronal network, we first examine the temporal evolution of the system in the low-noise regime. In this regime, the time-dependent synaptic coupling $\epsilon(t)$ continuously adjusts in response to the network activity, leading to alternating episodes of enhanced and reduced cluster coherence. Fig.\ref{fig:1} illustrates the characteristic \textit{synchronized cluster} phenomenon under low-noise conditions. The cluster error \(\delta x_{\text{clu}}(t)\) (panel (c)) oscillates rhythmically, with epochs of near-zero error (synchronized cluster) alternating with intervals of elevated error (desynchronized cluster). This breathing rhythm is tightly coupled to the astrocyte-mediated modulation of synaptic strength \(\epsilon(t)\) (panel (d)). When the cluster synchronizes, the resulting increase in the order parameter \(R(t)\) drives \(\epsilon(t)\) upward via Eq.~\ref{eq:astro}. This enhanced coupling eventually destabilizes the synchronized state, leading to desynchronization, which in turn reduces \(R(t)\) and allows \(\epsilon(t)\) to decay. Crucially, this behavior relies on the symmetric connectivity of the subset \(V_c\), which allows the astrocytic feedback to act uniformly on these specific nodes, driving the collective alternation. This feedback loop creates a self-sustained oscillation between synchronization and desynchronization.
The robustness of this breathing mechanism with respect to slower astrocytic timescales is further examined in Appendix~\ref{ap1}, where the same simulations are repeated by varying only the delay parameter \(\tau \in \{500, 1000, 1500\}\). The persistence of the synchronized--desynchronized alternations for larger delays confirms that the observed collective dynamics are not restricted to the default value of \(\tau=200\).

\begin{figure}
\centering
    \begin{tabular}{cc}
       (a) & (b)\\
       \includegraphics[width=0.235\textwidth]{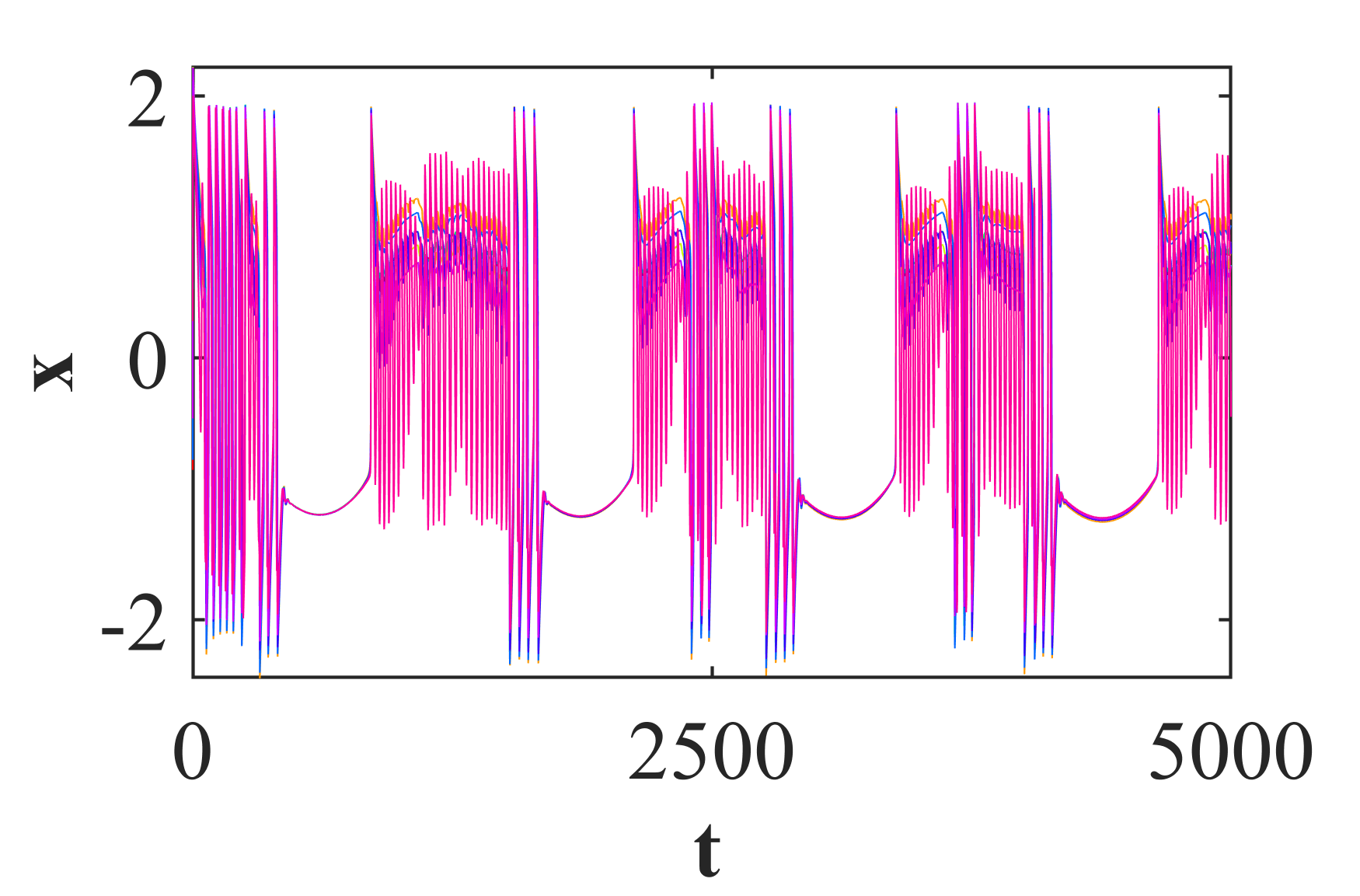} &  
       \includegraphics[width=0.235\textwidth]{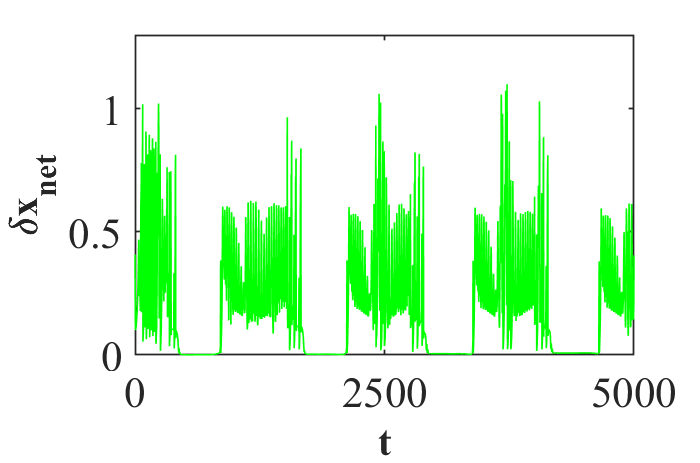}\\
       (c) & (d)\\
       \includegraphics[width=0.235\textwidth]{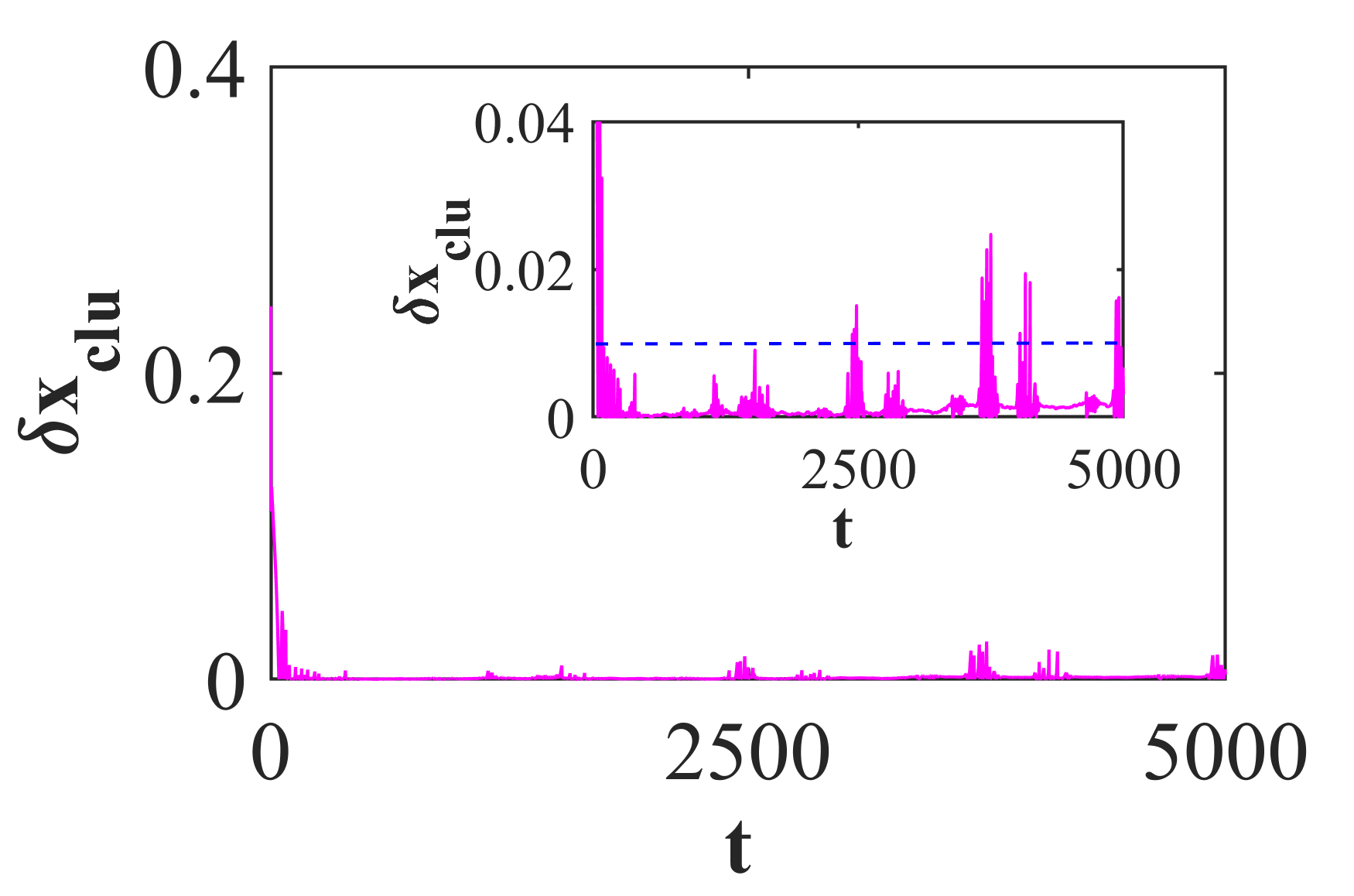}&
       \includegraphics[width=0.235\textwidth]{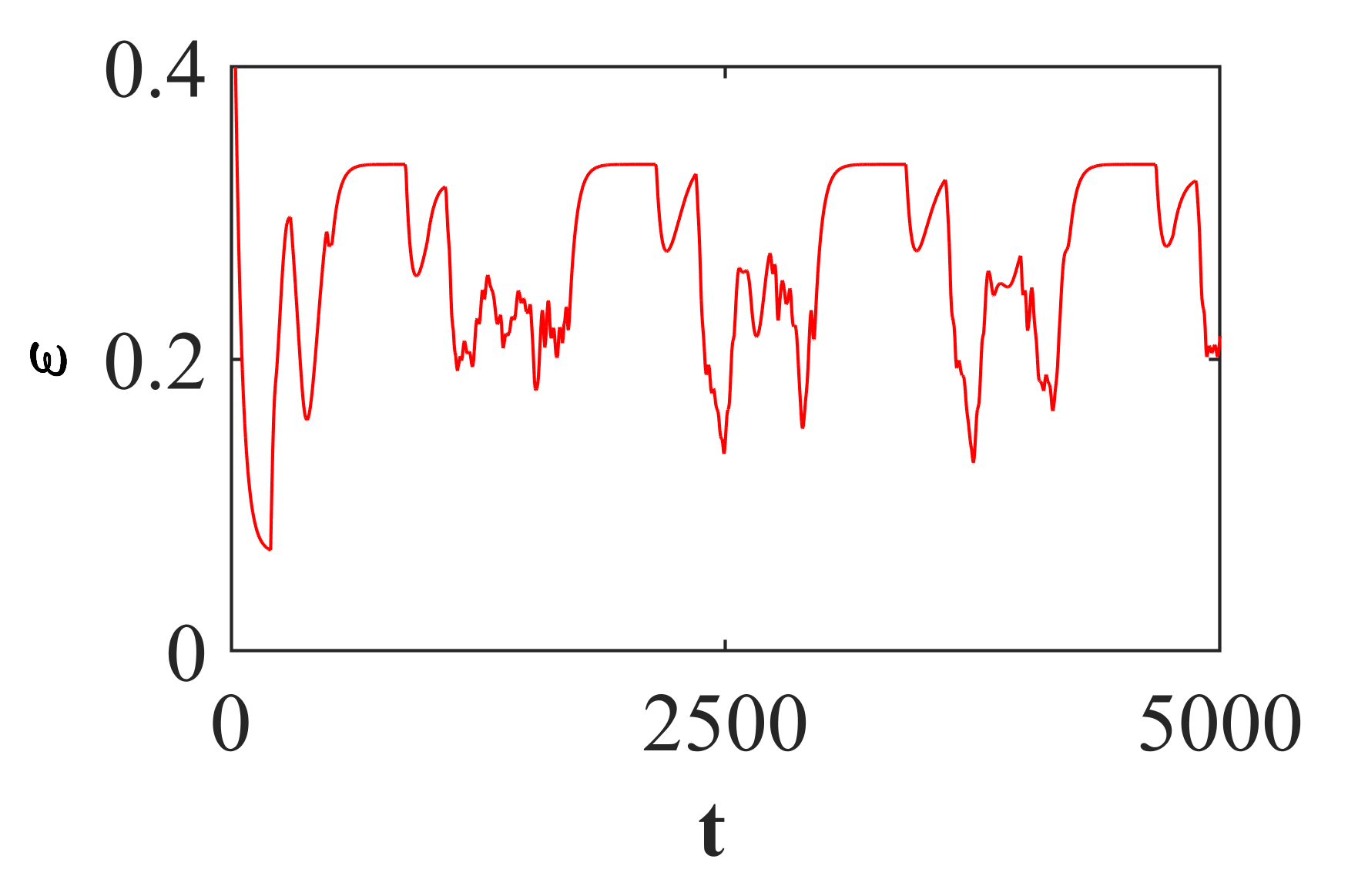}
    \end{tabular}
    \caption{Breathing dynamics under moderate noise (\(D = 0.1\)). (a) Membrane potentials \(x_i(t)\) of all neurons, showing clustered firing patterns. (b) Global synchronization error \(\delta x_{\text{net}}(t)\). (c) Cluster synchronization error \(\delta x_{\text{clu}}(t)\) for the symmetric neuron subset \(V_c = \{4,6,9\}\). The blue dashed line in the $\delta x_{\text{clu}}(t)$ panels indicates the desynchronization threshold $\delta_{\text{th}} = 0.01$. (d) Synaptic coupling strength \(\epsilon(t)\) modulated by astrocyte feedback. Parameters: \(\alpha = 0.03\), \(\beta = 0.008\), \(\gamma = 0.002\). The breathing rhythm persists but with increased irregularity in both amplitude and period.}
    \label{fig:2}
\end{figure}

Increasing noise to \(D = 0.1\) (see Fig.~\ref{fig:2}) enhances the breathing pattern by introducing greater variability in the oscillation amplitude and period. While the network does not achieve complete global synchronization (panel (b)), subsets of neurons, particularly those sharing symmetric connectivity, periodically transition between coherent and incoherent firing states (panel (c)). The underlying mechanism remains intact, yet the added stochastic perturbations reduce the regularity of the transitions, mirroring the noisier operational regime typical of biological neural circuits. This indicates that the astrocyte-neuron feedback loop possesses an inherent buffering capacity, absorbing moderate stochastic shocks without collapsing the macroscopic cluster organization. This robustness suggests that the breathing cluster phenomenon is not an artifact of deterministic dynamics but can persist under physiologically relevant levels of variability.

\subsection{Temperature Modulates Noise Sensitivity and Delay Robustness of Breathing Clusters}

\begin{figure}
    \centering
    \begin{tabular}{cc}
        (a) & (b) \\
        \includegraphics[width=0.235\textwidth]{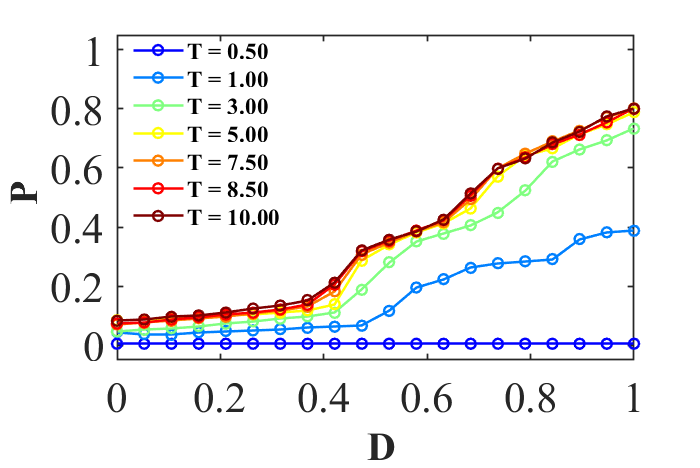} & 
        \includegraphics[width=0.235\textwidth]{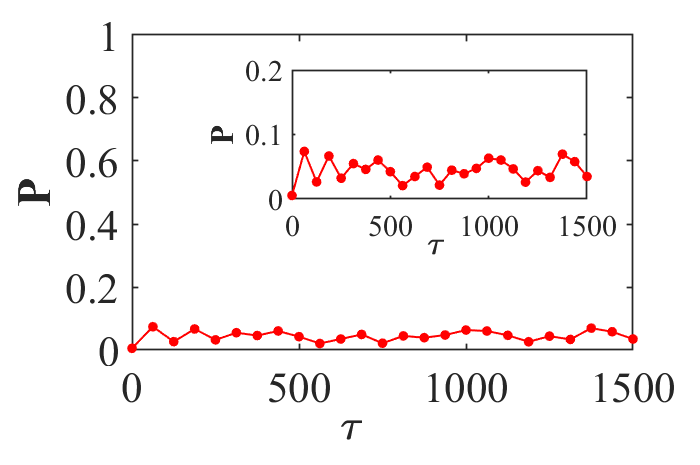}
    \end{tabular}
    \caption{Environmental tuning and delay robustness of breathing clusters. 
    (a) Desynchronization fraction $P$ as a function of noise amplitude $D$ for various temperatures $T$. 
    Lower temperatures ($T < 1$) confer high robustness ($P \approx 0$), while elevated $T$ shifts the stability boundary leftward, making the system susceptible to desynchronization at lower noise levels. 
    (b) Dependence of $P$ on the astrocytic feedback delay $\tau$. It shows that $P$ remains low across the investigated delay interval, indicating robustness of the breathing dynamics to astrocytic feedback delays. 
    Fixed parameters: $\alpha=0.03$, $\beta=0.008$, $\gamma=0.002$.}
    \label{fig:3}
\end{figure}

Figure~\ref{fig:3} quantifies how environmental factors and signaling kinetics reshape the functional boundaries of breathing clusters. 
Panel (a) reveals a strong temperature dependence of the desynchronization fraction $P$. 
At low temperatures ($T \lesssim 1$), the network exhibits remarkable robustness, with $P$ remaining near zero across the entire range of noise amplitudes, indicating that the breathing clusters maintain coherence even under strong stochastic perturbations. 
As $T$ increases, the curves shift progressively leftward; for instance, at physiological temperature ($T=5$), $P$ rises significantly as $D$ exceeds 0.4, whereas at $T=10$, desynchronization dominates ($P > 0.4$) even at moderate noise levels. 
This suggests that thermal activation alters the recovery kinetics (via the term $be^{1/T}$), effectively accelerating the neuronal timescale and amplifying the destabilizing impact of noise. 
Thus, temperature acts as a tunable control parameter that sets the noise threshold for synchronization breakdown.

Panel (b) examines the influence of the feedback delay $\tau$ in the astrocytic regulation (Eq.~5). The plot shows that the desynchronization fraction $P$ remains nearly constant for delays ranging from 0 to 1500. The investigated delay range, extending from the default value $\tau=200$ to $\tau=1500$, covers a substantial portion of the slower timescales associated with astrocytic calcium signaling and
gliotransmission ~\cite{haydon2006astrocyte,araque2014gliotransmitters,semyanov2020making,rusakov2025active}. The persistence of breathing dynamics across
this range indicates that the phenomenon does not require a narrowly
tuned astrocytic delay. To further verify that the characteristic breathing
dynamics remain preserved over slower astrocytic timescales, additional simulations are provided in Appendix~\ref{ap1}.
Nevertheless, astrocytic Ca$^{2+}$ responses can extend to longer
timescales, and exploring larger values of $\tau$ will be an important
direction for future work. 

\subsection{Astrocyte Parameters Govern Breathing Stability Across Noise Regimes}
 \begin{figure*}
    \centering
\includegraphics[width=1\textwidth]{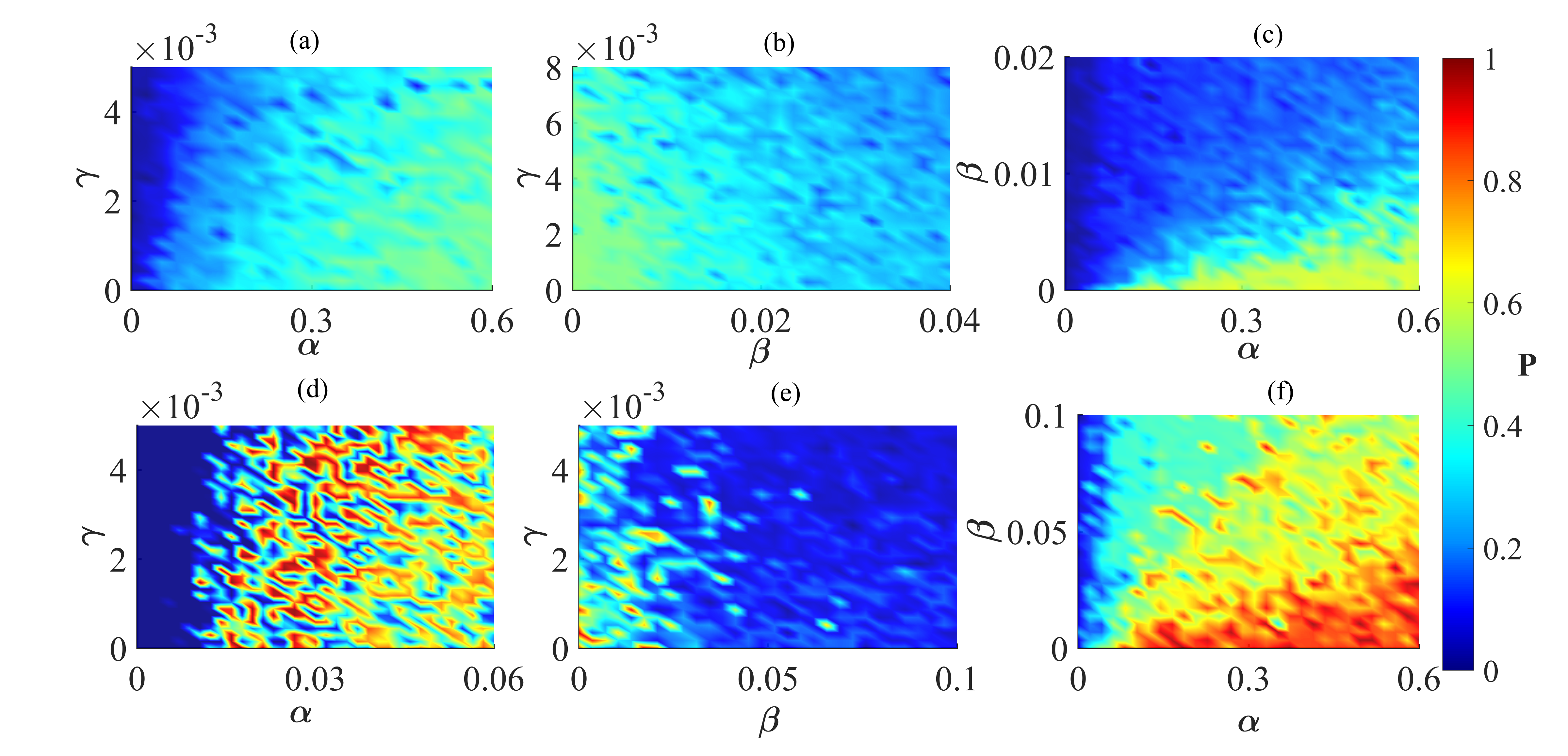}
    \caption{
Comparative parameter landscapes of breathing stability under quiescent and perturbed regimes. 
The top row corresponds to low-noise conditions ($D=0.01$), while the bottom row represents high-noise conditions ($D=1$). 
Each panel displays a two-dimensional map of the desynchronization fraction $P$: 
(a,d) in the $(\alpha,\gamma)$ plane with $\beta=0.008$, 
(b,e) in the $(\beta,\gamma)$ plane with $\alpha=0.05$, and 
(c,f) in the $(\alpha,\beta)$ plane with $\gamma=0.002$. 
Blue regions correspond to low desynchronization fractions ($P<0.2$), whereas yellow-to-red regions indicate increasing values of $P$. 
Because $P$ measures the fraction of time for which the cluster synchronization error exceeds the threshold, it does not by itself distinguish rigid synchronization from breathing states with a low desynchronized-time fraction. 
The temporal behavior of $\delta x_{\mathrm{clu}}(t)$ is therefore used to distinguish the different dynamical regimes. 
The two-dimensional landscapes reveal non-monotonic dependencies and parameter corridors associated with rhythmic breathing. 
Increased noise contracts and fragments the parameter regions supporting cluster switching, demanding more precise astrocytic tuning. 
Fixed parameters: $\tau=200$, $T=5$.
}
    \label{fig:4}
\end{figure*}
To characterize how astrocytic feedback shapes cluster dynamics under varying environmental conditions, we mapped the dependence of $P$ on key astrocyte parameters across low- and high-noise regimes (Fig.~\ref{fig:4}). Blue regions ($P<0.1$) denote stable synchronization, while yellow-to-red indicate intermittent breathing or desynchronization.

Under low noise ($D=0.01$, top row), the $(\alpha,\gamma)$ landscape (panel (a)) exhibits a diagonal corridor of rhythmic breathing, indicating that intermediate-to-high attenuation coefficients ($\alpha \approx 0.3$--$0.6$) combined with moderate basal glutamate release ($\gamma \approx 10^{-3}$ to $3\times10^{-3}$) optimally sustain intermittent switching. The $(\beta,\gamma)$ landscape (panel (b)) shows a non-monotonic dependence on $\gamma$ and an optimal intermediate band for $\beta$, confirming that basal drive and feedback gain act as bifurcation-like control parameters.

The $(\alpha,\beta)$ landscape (panel (c)) further demonstrates that the network's sensitivity to coupling strength $\beta$ is highly dependent on the decay rate $\alpha$. For intermediate $\alpha$, $P$ peaks at an optimal $\beta$ but drops to near zero (blue regions) when $\beta$ is either too weak or too strong. If $\alpha$ is too small, the parameter space is entirely blue, indicating that overly slow astrocytic decay locks the network into rigid synchronization regardless of $\beta$. This non-monotonic relationship reveals that astrocytes function as nonlinear regulators rather than linear amplifiers: weak coupling fails to overcome the network's natural tendency toward incoherence, while excessive coupling saturates the global synaptic strength, abolishing the intermittent transitions required for breathing.

Under high noise ($D=1$, bottom row), the parameter landscapes are dramatically reshaped. The breathing corridor in the $(\alpha,\gamma)$ plane contracts and fragments (panel (d)), with the viable $\alpha$ range narrowing significantly. This indicates that strong stochastic perturbations erode the robustness of rhythmic switching, demanding more precise tuning of astrocytic decay kinetics. The viable parameter regions in the $(\beta,\gamma)$ plane (panel (e)) also shrink and become highly irregular. Furthermore, the high-noise $(\alpha,\beta)$ map (panel (f)) reveals that elevated noise drastically reduces the system's tolerance for suboptimal feedback parameters, confining the breathing regime to a narrow 2D island of intermediate values rather than a continuous corridor. These comparative 2D maps demonstrate that while astrocytes can effectively tune neural cluster dynamics in quiescent regimes, their regulatory capacity operates closer to functional limits under noisy conditions, where small parameter deviations may trigger transitions to permanent desynchronization or rigid hypersynchrony.

\subsection{Modulation of Breathing Rhythm by Astrocyte Dynamics}

\begin{figure*}[htp]
\centering
\begin{tabular}{ccc}
(a)  & (b)  & (c)  \\
\includegraphics[width=0.29\textwidth]{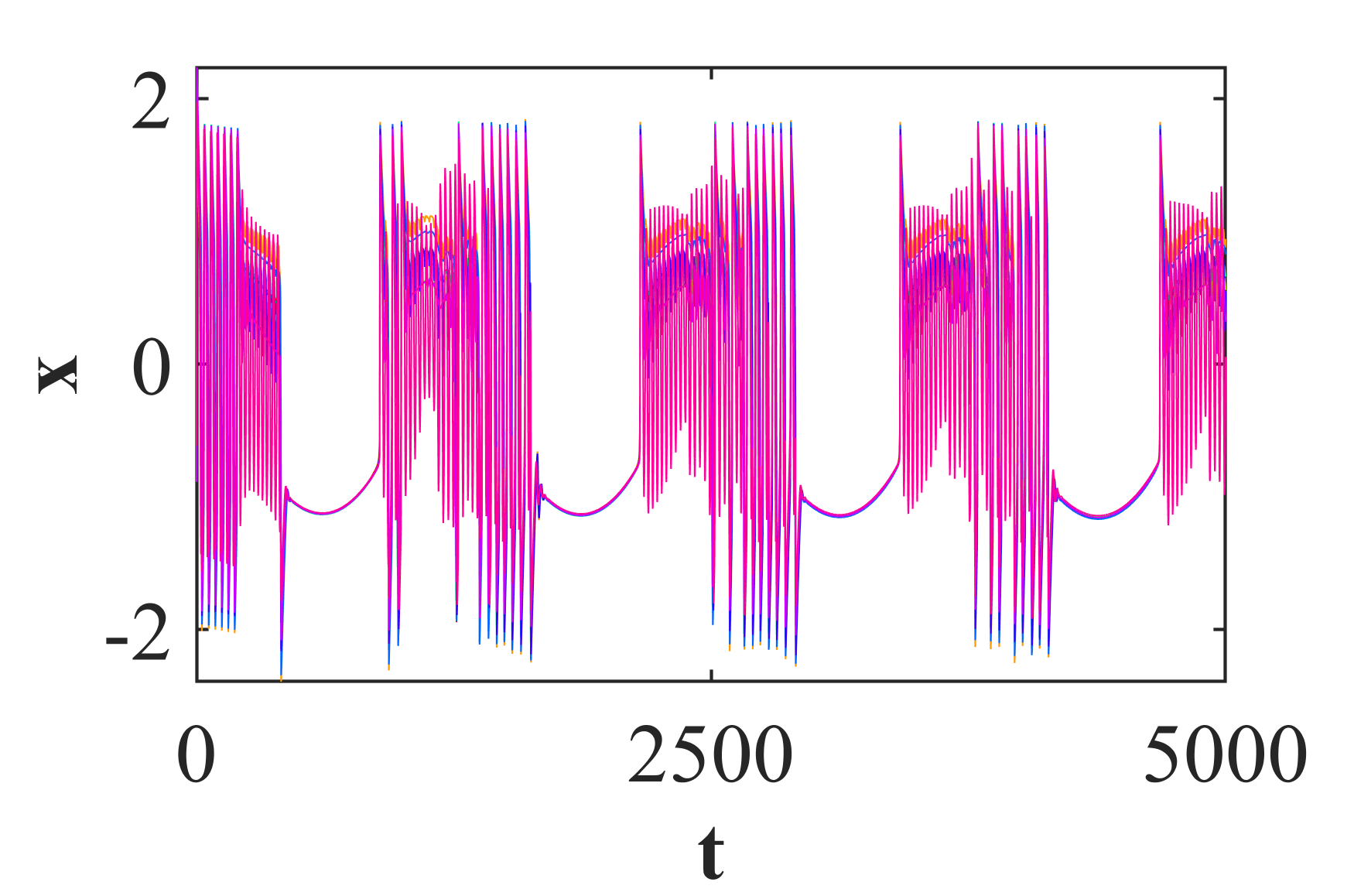} & 
\includegraphics[width=0.29\textwidth]{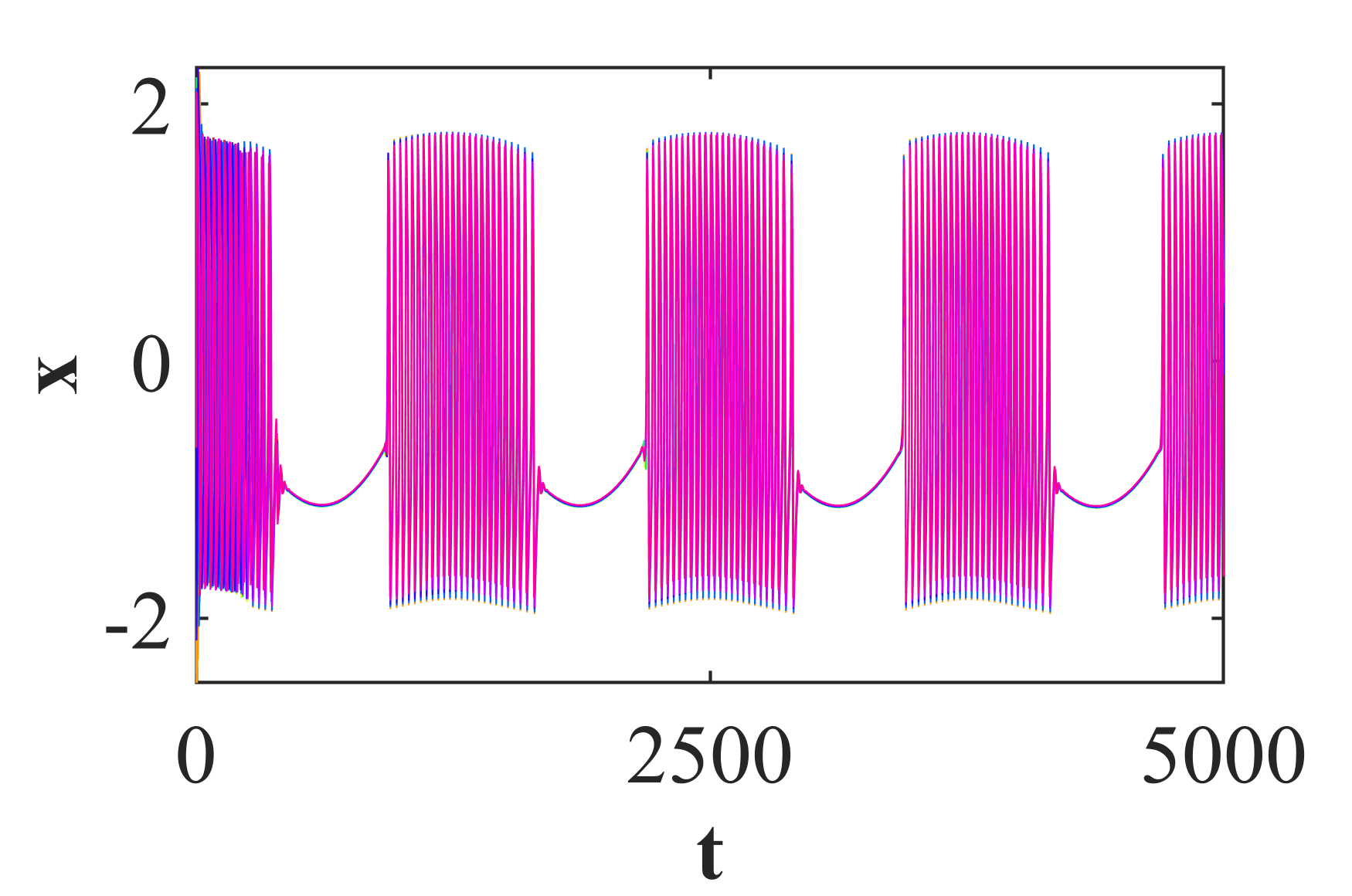} & 
\includegraphics[width=0.29\textwidth]{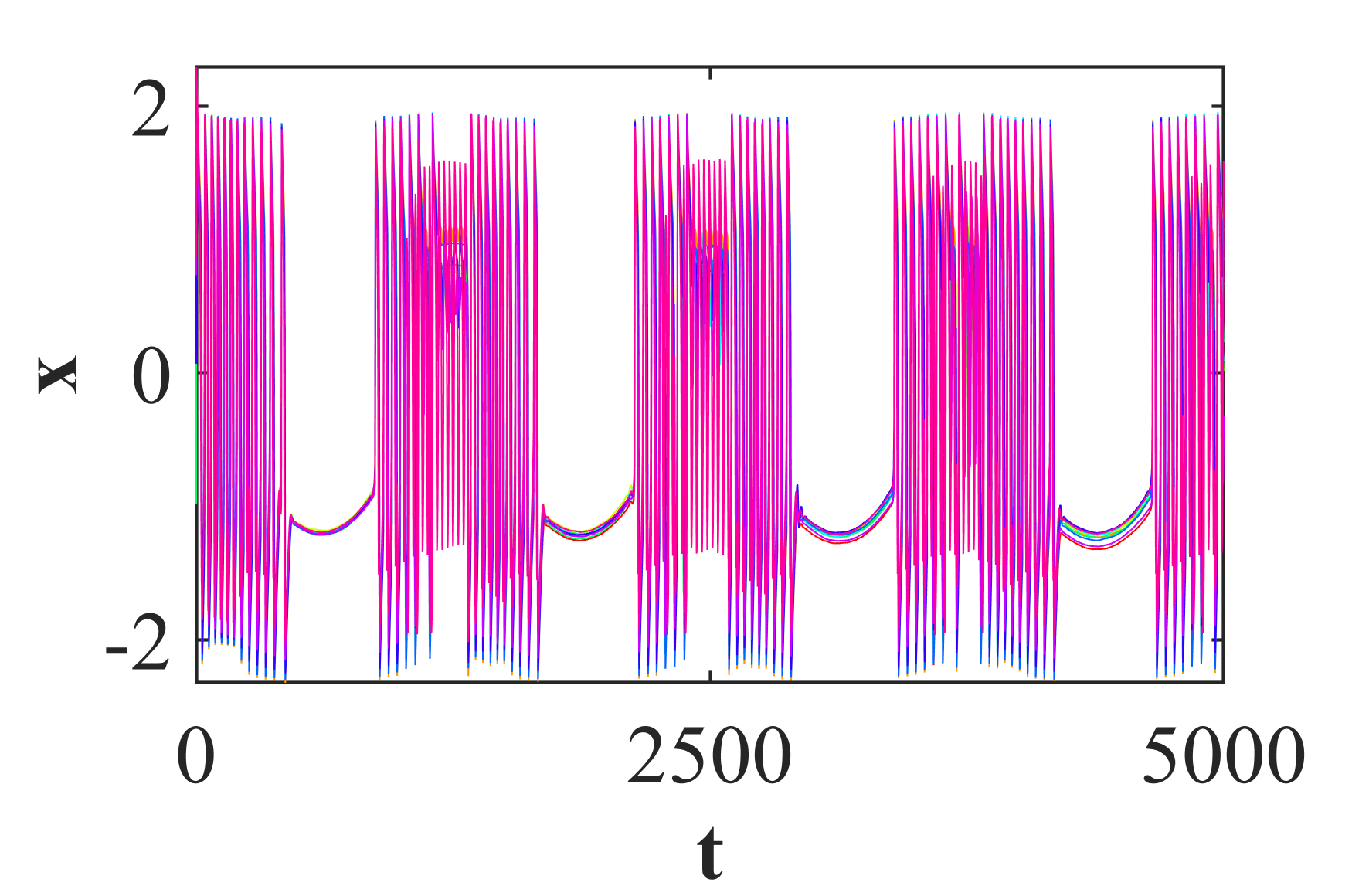} \\
(d)  & (e)  & (f)  \\
\includegraphics[width=0.29\textwidth]{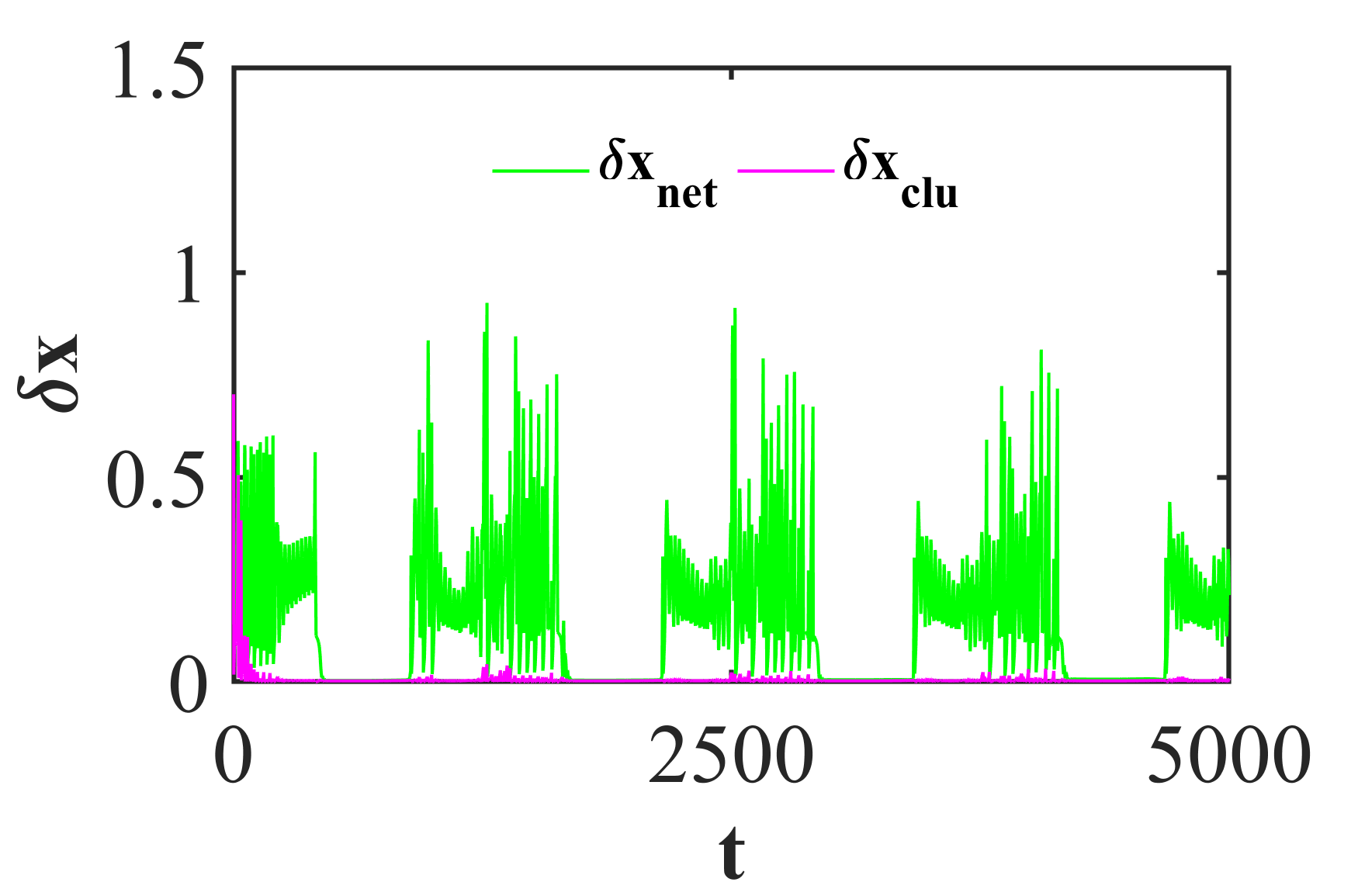} & 
\includegraphics[width=0.29\textwidth]{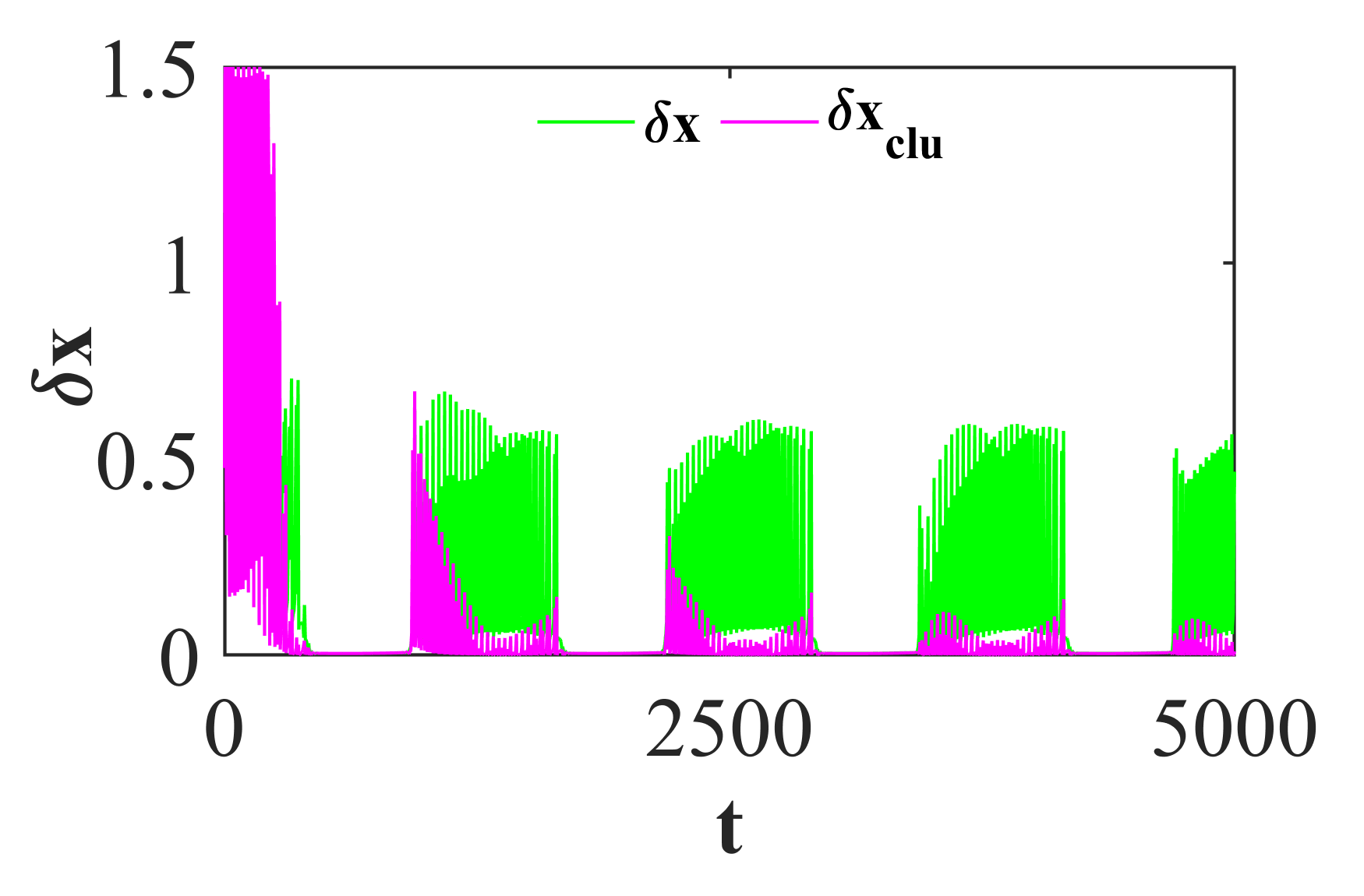} & 
\includegraphics[width=0.29\textwidth]{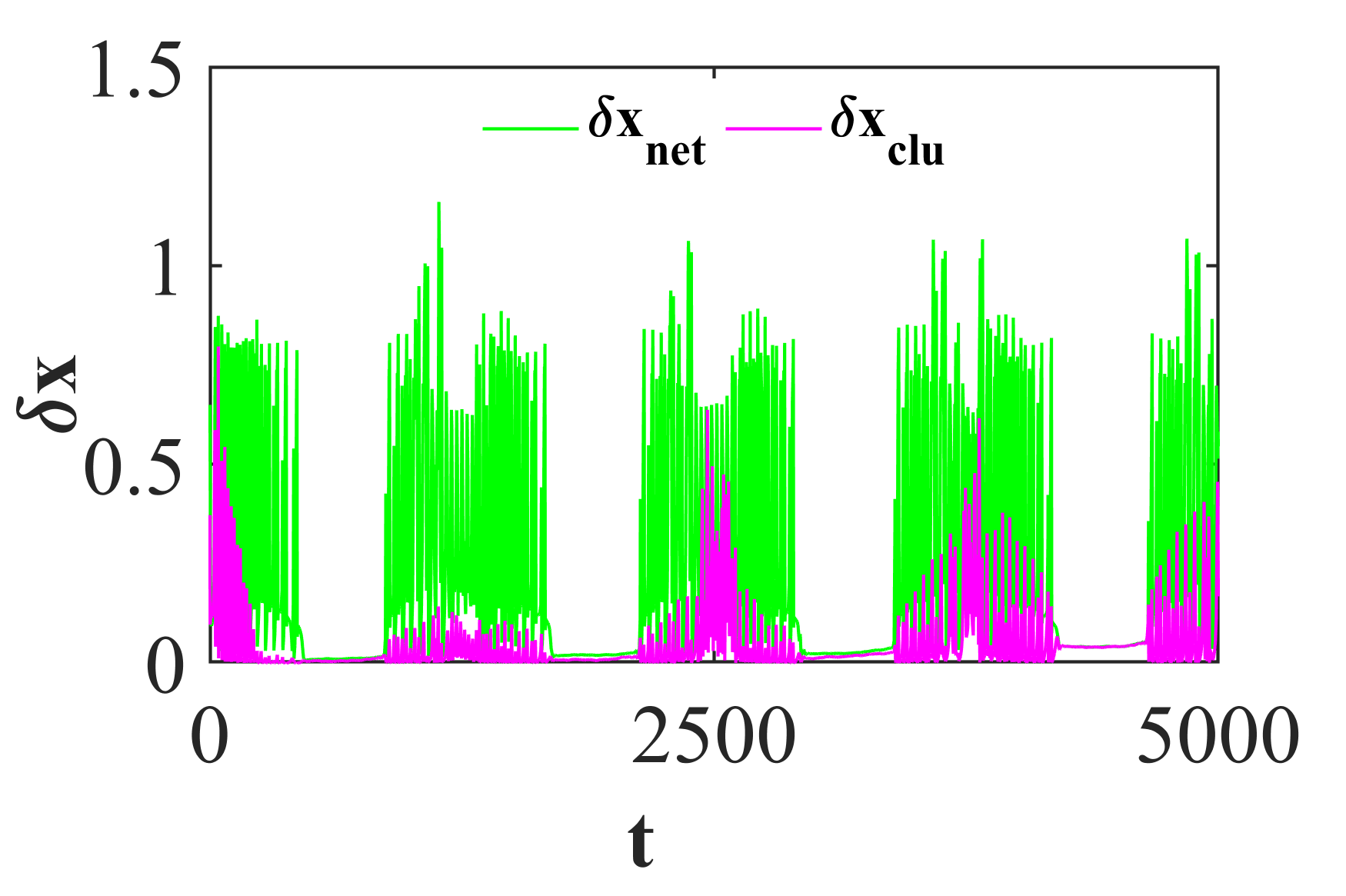} \\
\end{tabular}
\caption{Comparative analysis of breathing dynamics across three regimes. \textbf{Top row:} Membrane potentials $x_i(t)$ showing the spatial coherence of the clusters. \textbf{Bottom row:} Global synchronization error $\delta x_{\text{net}}(t)$ (green) and cluster synchronization error $\delta x_{\text{clu}}(t)$ (pink). \textbf{(a)} Baseline regime $\alpha=0.05$, $\gamma=0.005$,$\beta=0.008$ for $P=0.04$, low noise ($D=0.01$): stable breathing with slow astrocytic decay. \textbf{(b)} Fast decay regime $\alpha=0.3$, $\gamma=0.002$,$\beta=0.008$ for $P=0.4$, low noise ($D=0.01$): increasing $\alpha$ shortens the breathing period and reduces oscillation amplitude. \textbf{(c)} High noise regime ($D=1$), $\alpha=0.05$, $\gamma=0.002$, $\beta=0.008$ for $P=0.6$: strong stochastic perturbations disrupt rhythmic coherence and destabilize cluster boundaries.}
\label{fig:5}
\end{figure*}

Direct comparison of time traces in Fig.~\ref{fig:5} reveals how astrocyte properties and environmental variability sculpt the breathing rhythm. 

First, the role of astrocytic decay kinetics is evident when comparing panels (a) and (b). Increasing the attenuation coefficient $\alpha$ from $0.05$ to $0.3$ accelerates the decay of the astrocytic influence $\epsilon(t)$, which in turn shortens the breathing period and reduces the amplitude of $\delta x_{\text{clu}}(t)$ oscillations. This faster feedback prevents the cluster from settling into prolonged synchronized or desynchronized states, producing a rapid, shallow switching pattern. These timescales overlap with the temporal structure of cortical state-switching observed during attentional tasks~\cite{buzsaki2004neuronal,steriade2001natural}, suggesting that astrocytes could serve as tunable controllers of network timescales, enabling the brain to adapt its dynamical repertoire to varying computational demands~\cite{poskanzer2011astrocytic,poskanzer2016astrocytes}.

Second, the impact of stochastic perturbations is illustrated in panel (c). Under high noise conditions ($D=1$), the breathing rhythm becomes markedly irregular. Cluster boundaries are less distinct, and transitions between states lose their periodic regularity. This degradation underscores the challenge of maintaining coordinated dynamics in noisy regimes and emphasizes the importance of astrocyte-mediated stabilization mechanisms for preserving functional cluster organization in vivo.

Together, these results demonstrate that while astrocyte parameters can fine-tune the frequency of cluster switching to match cognitive requirements, there is a fundamental limit to this regulation when external noise exceeds the system's buffering capacity. These observations set the stage for understanding how pathological disruptions in glial homeostasis might push the network out of this functional buffering zone, leading to the rigid or chaotic states observed in neurological disorders.

\section{Discussion}
\label{sec:discussion}

The central result of this study is that breathing cluster synchronization persists as a structurally stable collective phenomenon when neuron--astrocyte networks are extended to include thermosensitive neuronal dynamics, intrinsic electric-field interactions, delayed astrocytic feedback, and stochastic perturbations. As illustrated by the persistent alternation between coherent and incoherent cluster states in Figs.~\ref{fig:1} and~\ref{fig:2}, the breathing mechanism is not a fragile artifact of deterministic or simplified modeling assumptions. Consistent with the symmetry-based framework proposed by Wang et al.~\cite{wang2023breathing}, the emergence of breathing clusters remains fundamentally governed by the presence of structurally symmetric nodes and astrocyte-mediated feedback. The electric-field variable was introduced primarily to assess the robustness of breathing synchronization under ephaptic-like interactions rather than to serve as the primary mechanism generating cluster breathing. While electric fields are known to influence neuronal excitability and collective dynamics through nonsynaptic interactions~\cite{bikson2004effects,radman2007spike,simo2021chimera}, they do not alter the underlying topological mechanism responsible for cluster formation. These findings support the view that breathing synchronization represents a robust network property that persists under additional biophysical influences.

A central contribution of this work is the non-monotonic dependence of breathing dynamics on the astrocytic coupling strength $\beta$ and attenuation coefficient $\alpha$. The parameter landscapes shown in Fig.~\ref{fig:4} reveal that breathing emerges only within an intermediate corridor of astrocytic regulation. When $\beta$ is too weak, astrocytic feedback fails to sustain rhythmic alternation between synchronized and desynchronized states. Conversely, excessively large $\beta$ saturates the synaptic strength $\varepsilon(t)$ and drives the network toward rigid synchronization. Similarly, overly small values of $\alpha$ prevent the system from escaping synchronized states. Only intermediate values of $\alpha$ and $\beta$ support robust breathing dynamics. The two parameters play distinct dynamical roles: $\beta$ acts as a regime switch determining whether breathing exists, whereas $\alpha$ governs how breathing occurs by regulating its period and amplitude. As shown in Fig.~\ref{fig:5}, increasing $\alpha$ shortens the breathing period and reduces oscillation amplitude without eliminating the breathing state itself. In other words, $\beta$ decides whether the system breathes, whereas $\alpha$ determines how fast and how deeply it breathes. This separation of functions highlights astrocytes as nonlinear regulators capable of controlling collective network dynamics independently of intrinsic neuronal firing properties, consistent with experimental evidence demonstrating bidirectional astrocytic modulation of synaptic transmission and network excitability~\cite{haydon2006astrocyte,tang2017astrocyte}. Recent computational studies have similarly shown that astrocytic feedback can regulate synchronization transitions and firing activity in heterogeneous neuronal networks~\cite{chen2024astrocyte,chen2025neural}.

Another notable result is the remarkable robustness of breathing clusters to delayed astrocytic feedback. As shown in Fig.~\ref{fig:3} panel(b), stable rhythmic alternation persists for delays ranging from $0$ to $1500$. Mechanistically, this insensitivity to precise timing arises because breathing is governed by a global feedback loop in which the coherence measure $R(t)$ integrates the activity of the entire network before modulating $\varepsilon(t)$. Delays shift the astrocytic response in time but leave the overall feedback architecture intact, thereby preserving the alternation between synchronized and desynchronized states. Such delay tolerance is consistent with the relatively slow timescales of astrocytic calcium signaling and gliotransmission~\cite{haydon2006astrocyte,perea2007astrocytes}. More generally, these findings suggest that breathing synchronization depends primarily on the feedback structure itself rather than on precise temporal alignment between neuronal firing and astrocytic responses.

In contrast, stochastic perturbations substantially constrain the parameter region supporting breathing dynamics. Although moderate noise preserves rhythmic alternation (Fig.~\ref{fig:2}), increasing noise progressively destabilizes breathing and fragments the viable parameter corridors. This effect is reflected in the behavior of the desynchronization fraction $P$ shown in Fig.~\ref{fig:3} panel(a), as well as in the parameter maps of Fig.~\ref{fig:4}, where the breathing regions contract markedly under elevated noise levels. 
To complement the phase-diagram analysis, Appendix.~\ref{ap2} provides representative time series illustrating the distinct dynamical regimes identified from the temporal evolution of the cluster synchronization error $\delta x_{\mathrm{clu}}(t)$. In particular, these examples clarify the distinction between rigid synchronization, breathing synchronization, and persistent desynchronization, which may not be uniquely captured by the desynchronization fraction $P$. While $P$ measures the fraction of time for which $\delta x_{\mathrm{clu}}(t)$ exceeds the desynchronization threshold, the temporal alternation between synchronized and desynchronized states provides the characteristic signature of breathing dynamics.
 
Under strong perturbations, only a narrow range of astrocytic parameters sustains breathing, while small deviations drive the system toward either persistent desynchronization or rigid synchrony. Temperature further modulates this behavior by shifting the noise threshold at which synchronization breakdown occurs. Together, these results indicate that astrocytic feedback provides a buffering mechanism against physiological variability but operates within finite functional limits under highly perturbed conditions, consistent with previous studies of noise-induced transitions in neural systems~\cite{zhu2021effects,nowotny2007dynamical}. The contraction of the breathing regime under strong perturbations suggests that astrocytic regulation becomes increasingly sensitive to parameter fluctuations as environmental variability increases.

From a biological perspective, the persistence of breathing clusters under thermosensitive dynamics, electric-field interactions, delays, and noise suggests that the phenomenon may represent a general organizing principle of astrocyte-regulated neural networks rather than a model-specific artifact. The observed transitions between coherent and incoherent activity may provide qualitative insight into mechanisms that balance network stability and flexibility. More broadly, these findings support the emerging view that astrocytes actively participate in shaping large-scale neural dynamics and cortical state transitions~\cite{poskanzer2011astrocytic,poskanzer2016astrocytes,steriade2001natural,buzsaki2004neuronal}. Recent theoretical studies further suggest that astrocytes can drive transitions between distinct dynamical regimes and reshape network activity patterns through slow regulatory feedback mechanisms~\cite{jacquir2025quiescence}. The ability of astrocytic feedback to regulate the timing and amplitude of breathing further suggests that glial cells may function as adaptive modulators of collective neural activity across multiple temporal scales. Recent reviews have further emphasized the role of astrocytes as active participants in circuit modulation through calcium-dependent signaling and gliotransmission~\cite{rusakov2025active}.

Several limitations should be acknowledged. Following Wang et al.~\cite{wang2023breathing}, the astrocytic dynamics were represented by a simplified linear model with} a single astrocyte coupled globally to the network. While this choice allows us to isolate the core feedback mechanism, it neglects the spatially restricted domains of individual astrocytes observed in biological tissue, where each astrocyte modulates a local subset of synapses~\cite{postnov2009dynamical,gordleeva2021modeling,cresswell2018compartmental}. This approach also neglects spatial heterogeneity, calcium-wave propagation, and the diversity of astrocytic populations observed in biological tissue. Furthermore, while exact network symmetry determines cluster composition in the present framework, biological circuits are more likely to exhibit approximate or dynamically evolving symmetries. The network size considered here ($N=10$) also remains small compared with cortical populations. Whether breathing synchronization persists in larger networks with heterogeneous connectivity and approximate symmetries therefore remains an open question. Future work should incorporate nonlinear calcium dynamics, multiple interacting astrocytes with local coupling domains, and more realistic network architectures to further assess the generality of the phenomenon~\cite{postnov2009dynamical,gordleeva2021modeling}.

Taken together, these results establish breathing cluster synchronization as a robust consequence of the interplay between network symmetry and astrocyte-mediated feedback. By demonstrating its persistence under thermosensitive excitability, intrinsic electric-field coupling, delayed feedback, and physiological noise, the present work substantially extends the applicability of the original breathing-cluster framework proposed by Wang et al.~\cite{wang2023breathing}. More generally, the results reinforce the concept of astrocytes as adaptive regulators of collective neural dynamics and provide a foundation for future investigations into the role of glial feedback in shaping large-scale brain activity.

\section{Conclusion}

This study demonstrates that breathing cluster synchronization remains a robust collective phenomenon in neuron--astrocyte networks incorporating thermosensitive neuronal dynamics, intrinsic electric-field interactions, delayed astrocytic feedback, and stochastic perturbations. Consistent with the symmetry-based mechanism underlying cluster formation, breathing dynamics persist across a broad range of biophysical conditions and are primarily regulated by astrocytic feedback. We show that astrocytic coupling strength and attenuation rate exert distinct dynamical roles, with $\beta$ determining the existence of breathing states and $\alpha$ controlling their temporal characteristics. Furthermore, breathing synchronization exhibits remarkable tolerance to delayed feedback, while elevated noise progressively restricts the parameter region supporting rhythmic alternation. Together, these results establish breathing clusters as a robust consequence of the interplay between network symmetry and astrocyte-mediated regulation.

More broadly, our findings reinforce the emerging view that astrocytes act as active modulators of large-scale neural dynamics rather than passive support cells. Future work should incorporate nonlinear calcium signaling, multiple interacting astrocytes, and heterogeneous network architectures to evaluate the generality of the proposed mechanism under more realistic physiological conditions.

\begin{acknowledgments}
ELFN  acknowledges financial support from CAPES (Grant No.~0001).
The authors FFF and TN gratefully acknowledges the financial support by National Institute of Science and Technology in Innovative Research in Health Sciences – from Nanotechnology to Artificial Intelligence (INCT PICS) sponsored by Brazil’s National Council for Scientific and Technological Development (CNPq), grant no. 408417/2024-2 and São Paulo Research Foundation (FAPESP), grant no. 2025/26818-7.Also, FFF and TN acknowledge the financial support by FAPESP grant no. 2025/18142-3, FFF thanks CNPq grants no.311989/2025-0. 
\end{acknowledgments}

\section*{Conflict of interest}

The authors declare that they have no conflict of interest.
\section*{Data availability}

The data that support the findings of this study are available
from the corresponding author upon reasonable request.

\appendix
\section{Robustness of the breathing dynamics to astrocytic delay}\label{ap1}
\begin{figure*}[htp!]
    \centering
    \begin{tabular}{cccc}
       (a) & (b) & (c) & (d)\\     \includegraphics[width=0.235\textwidth]{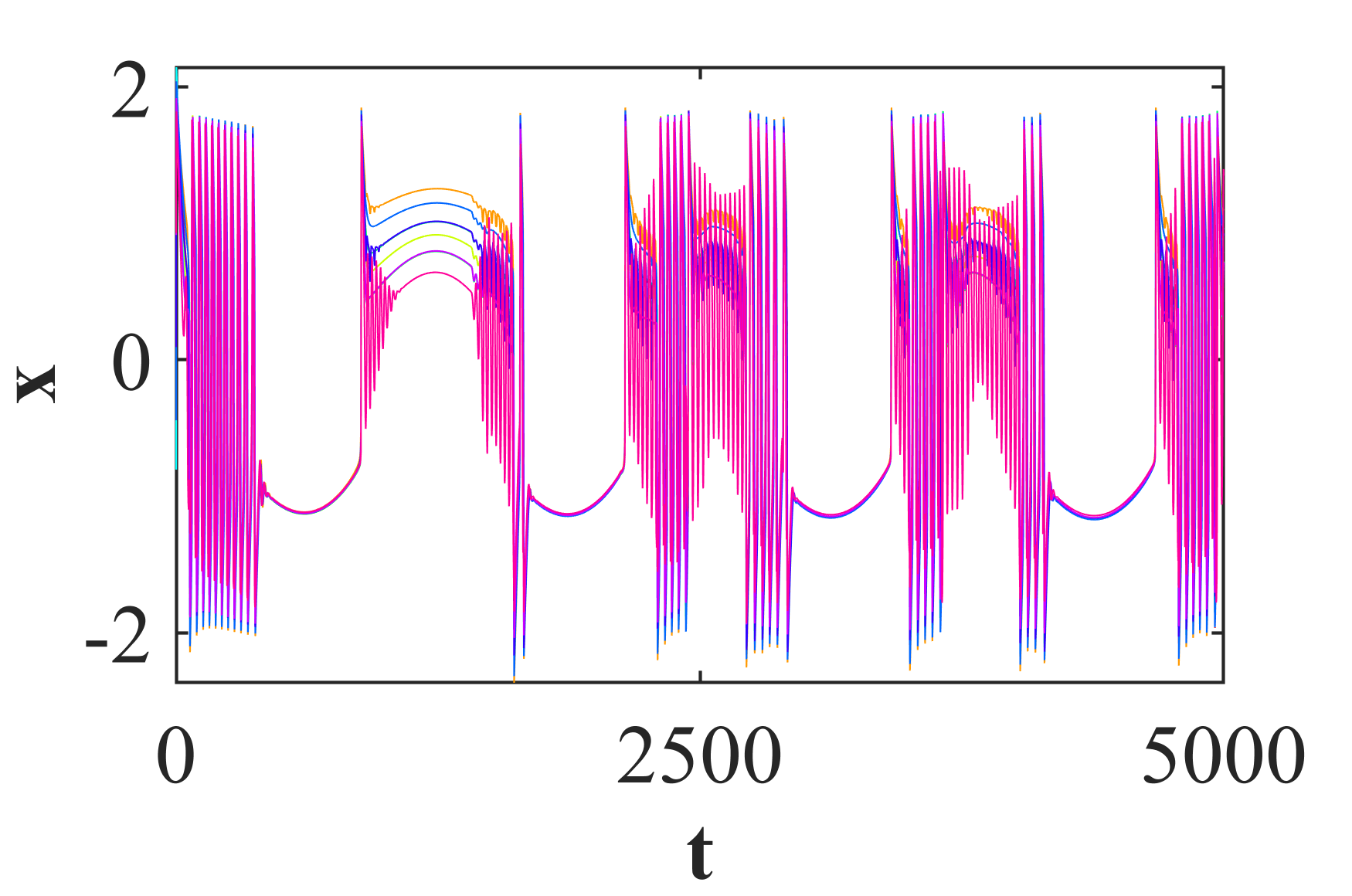} & \includegraphics[width=0.235\textwidth]{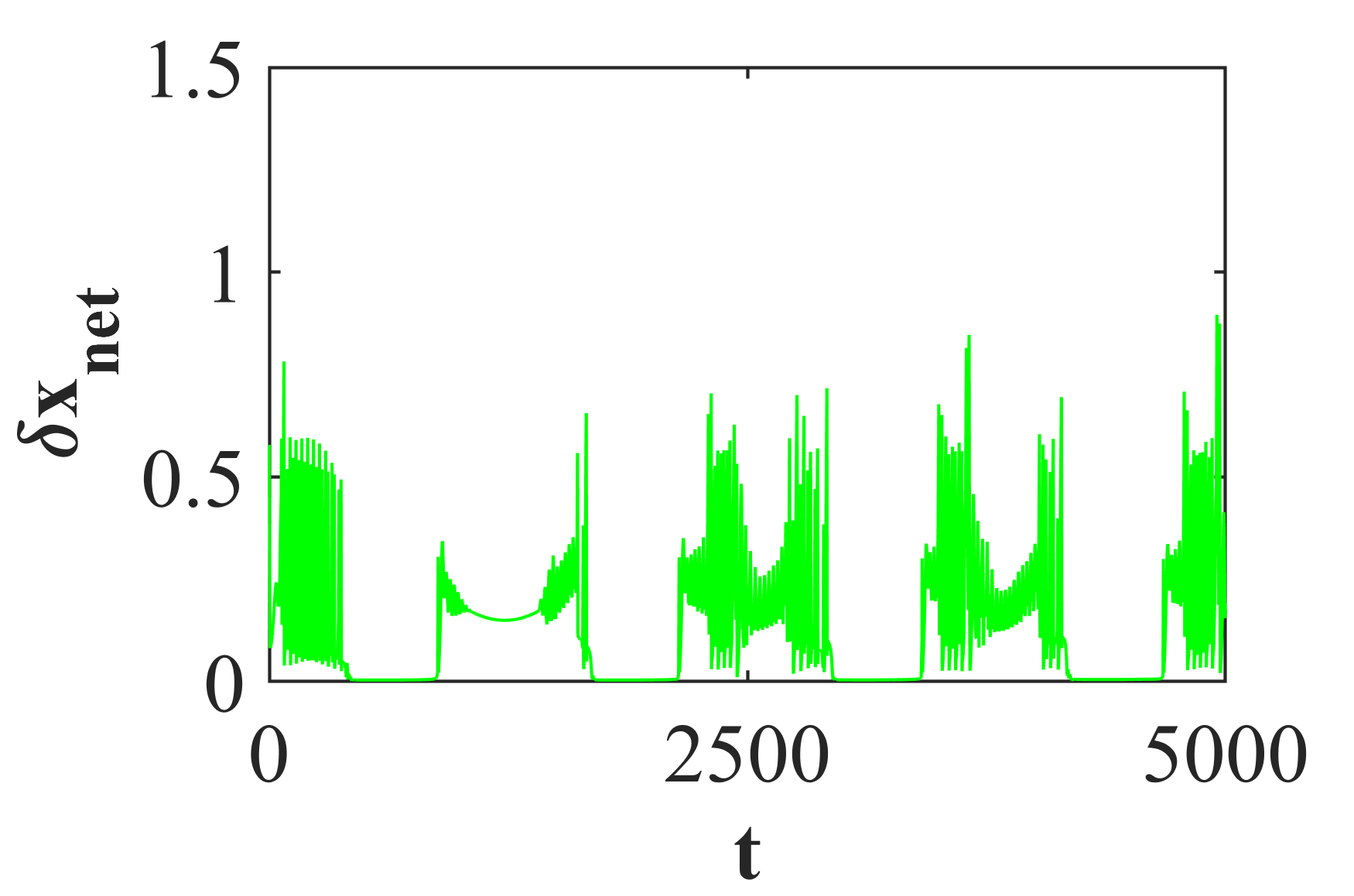}& \includegraphics[width=0.235\textwidth]{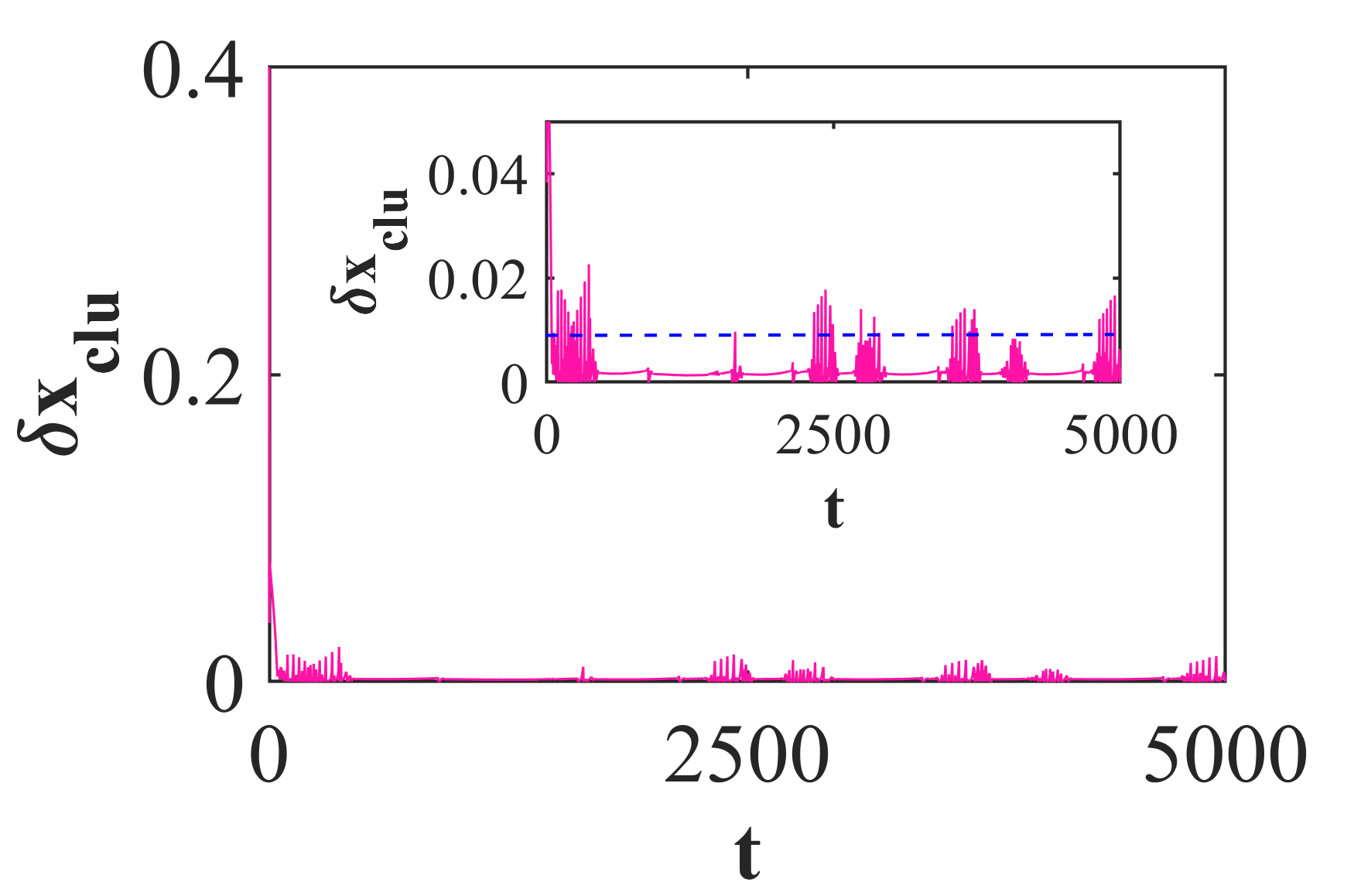} &        \includegraphics[width=0.235\textwidth]{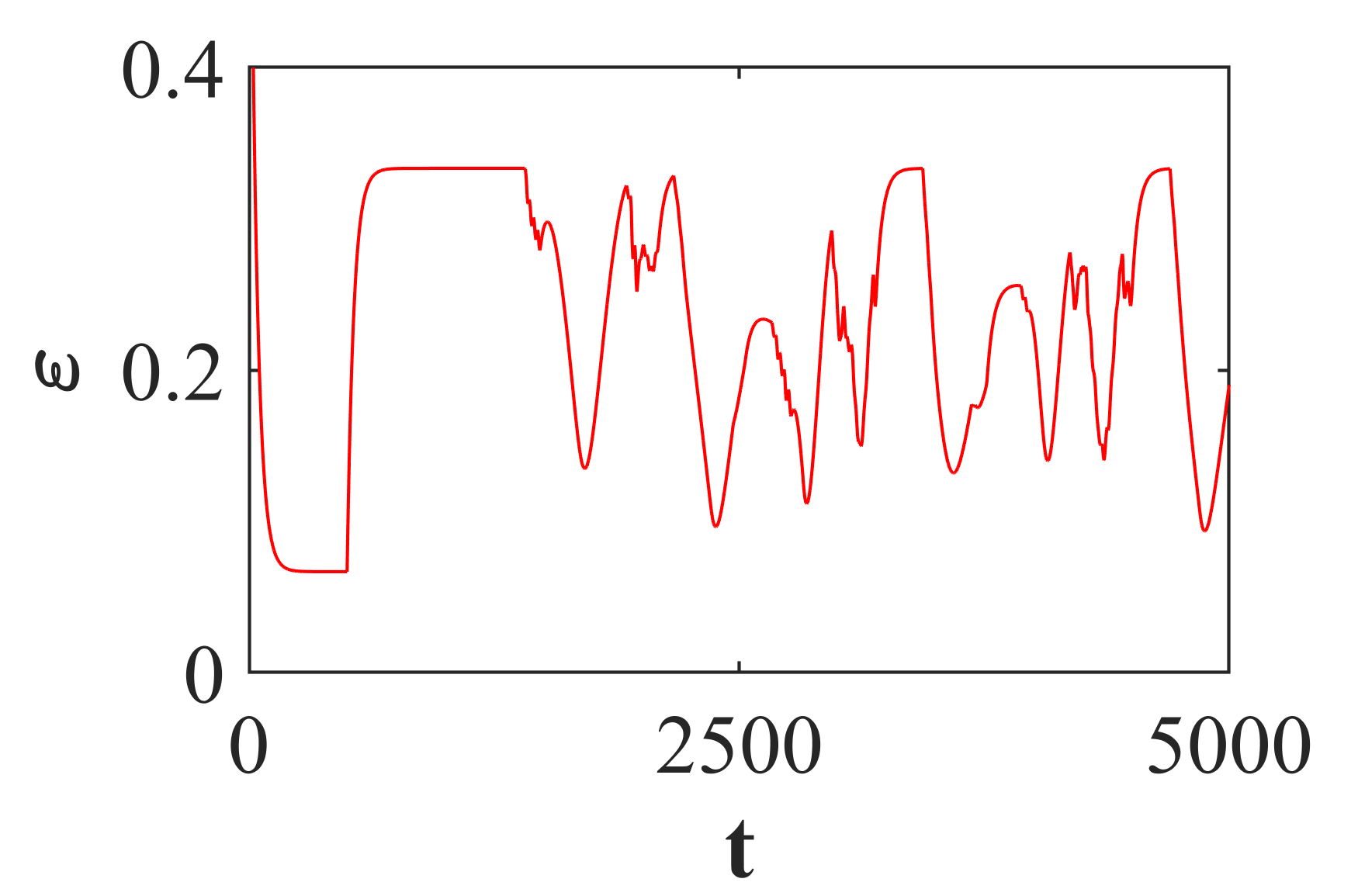}\\
       (e) & (f) & (g) & (h)\\       \includegraphics[width=0.235\textwidth]{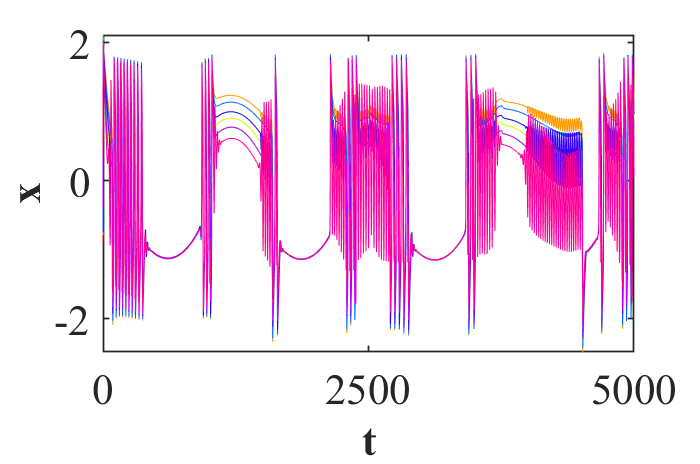}& \includegraphics[width=0.235\textwidth]{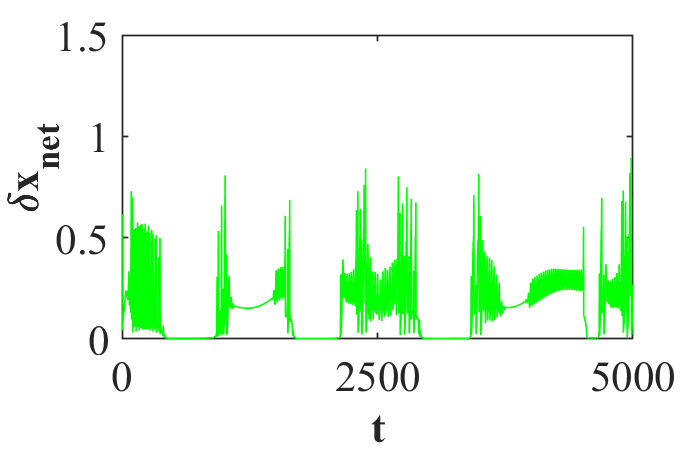} & \includegraphics[width=0.235\textwidth]{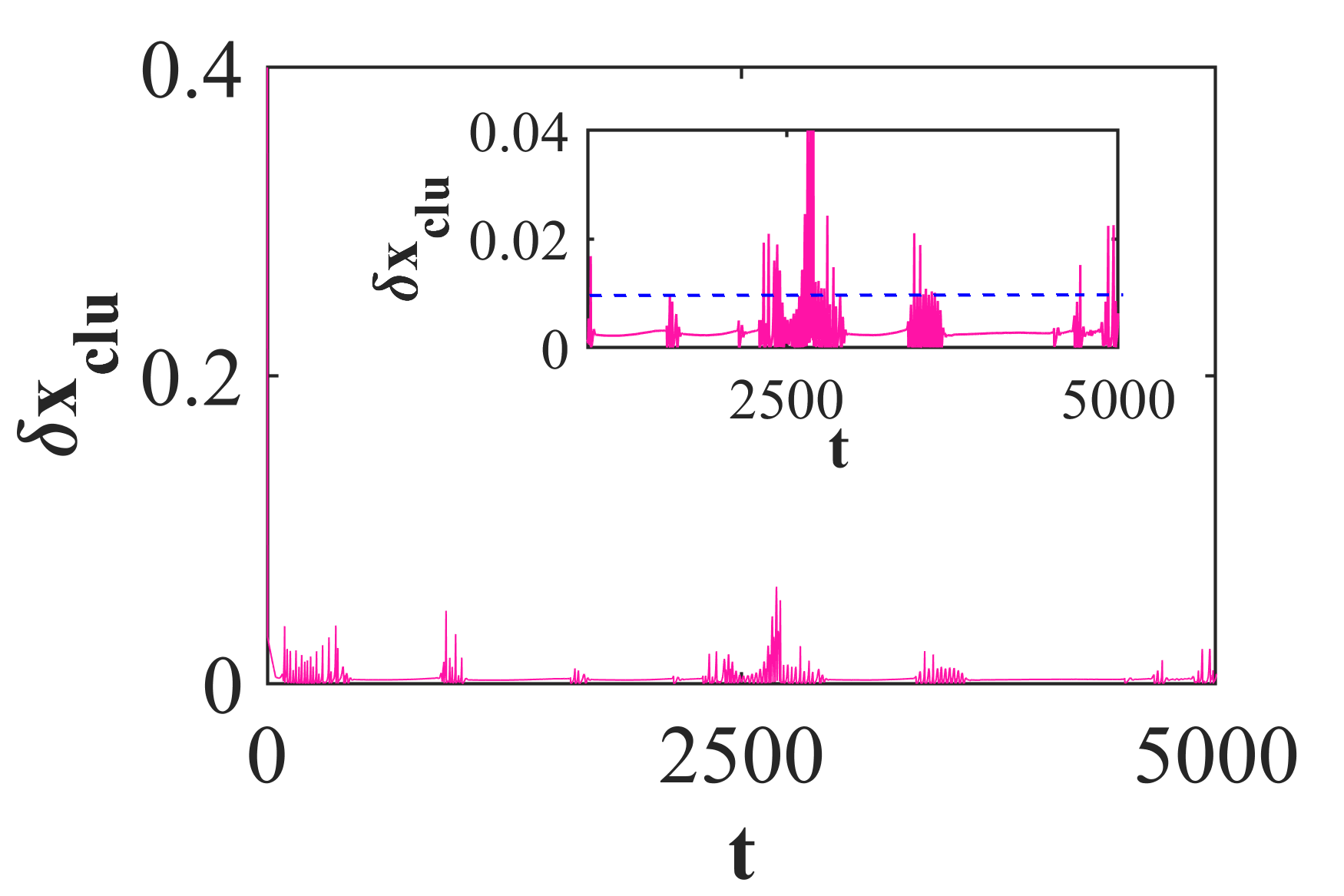} &     \includegraphics[width=0.235\textwidth]{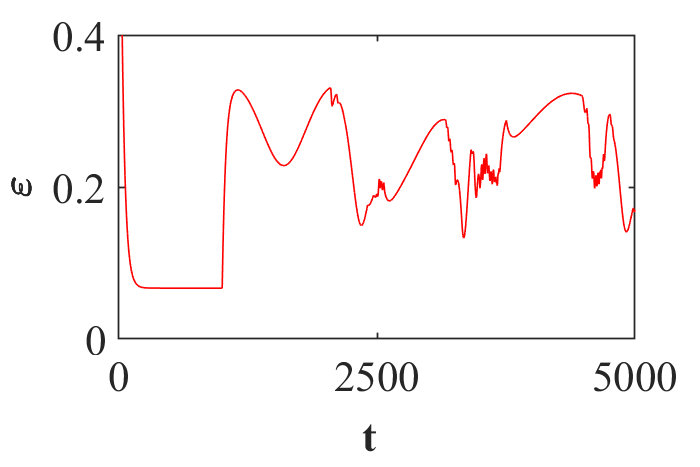}\\
     (i) & (j) & (k) & (l)\\
     \includegraphics[width=0.235\textwidth]{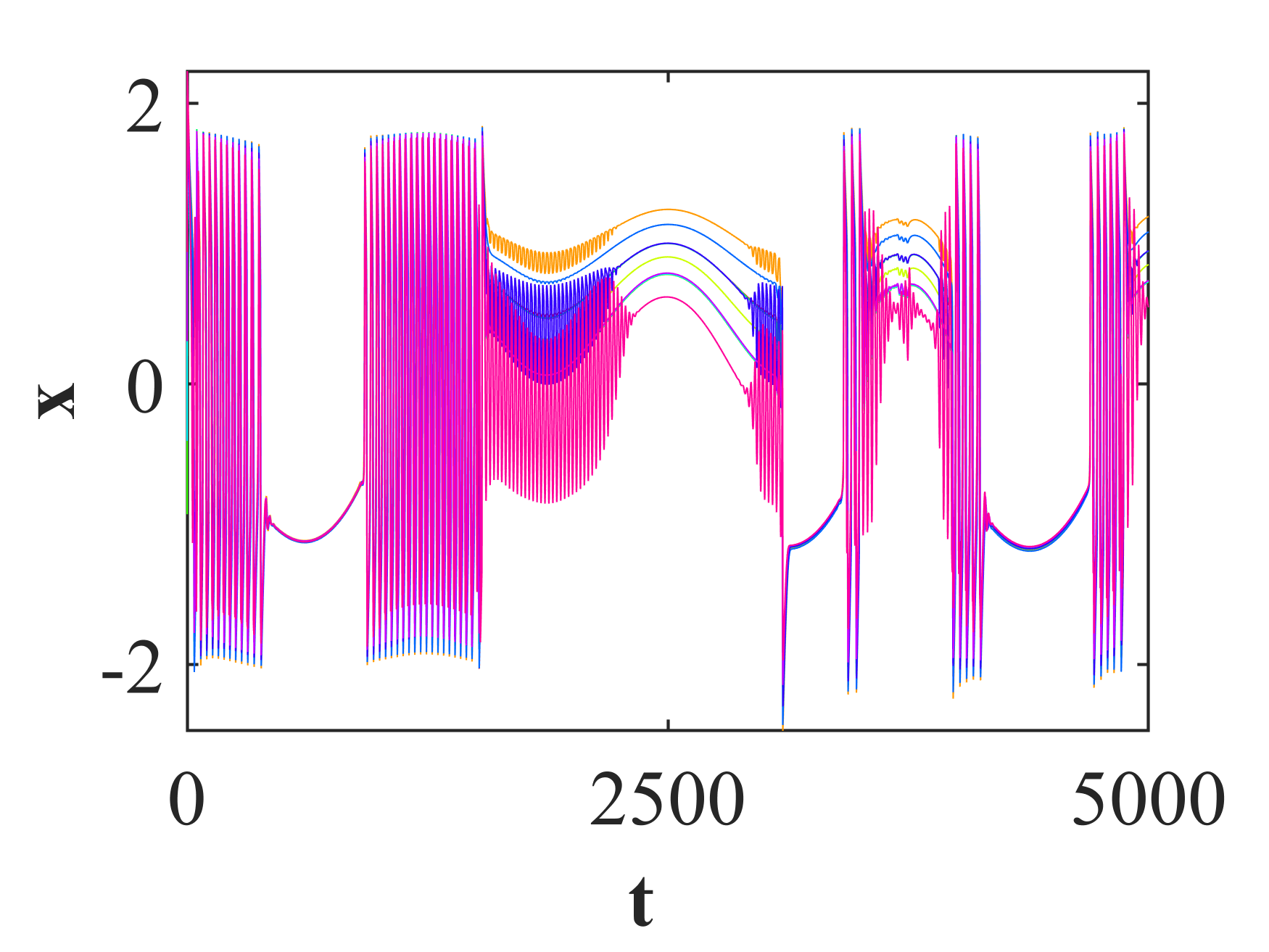}& \includegraphics[width=0.235\textwidth]{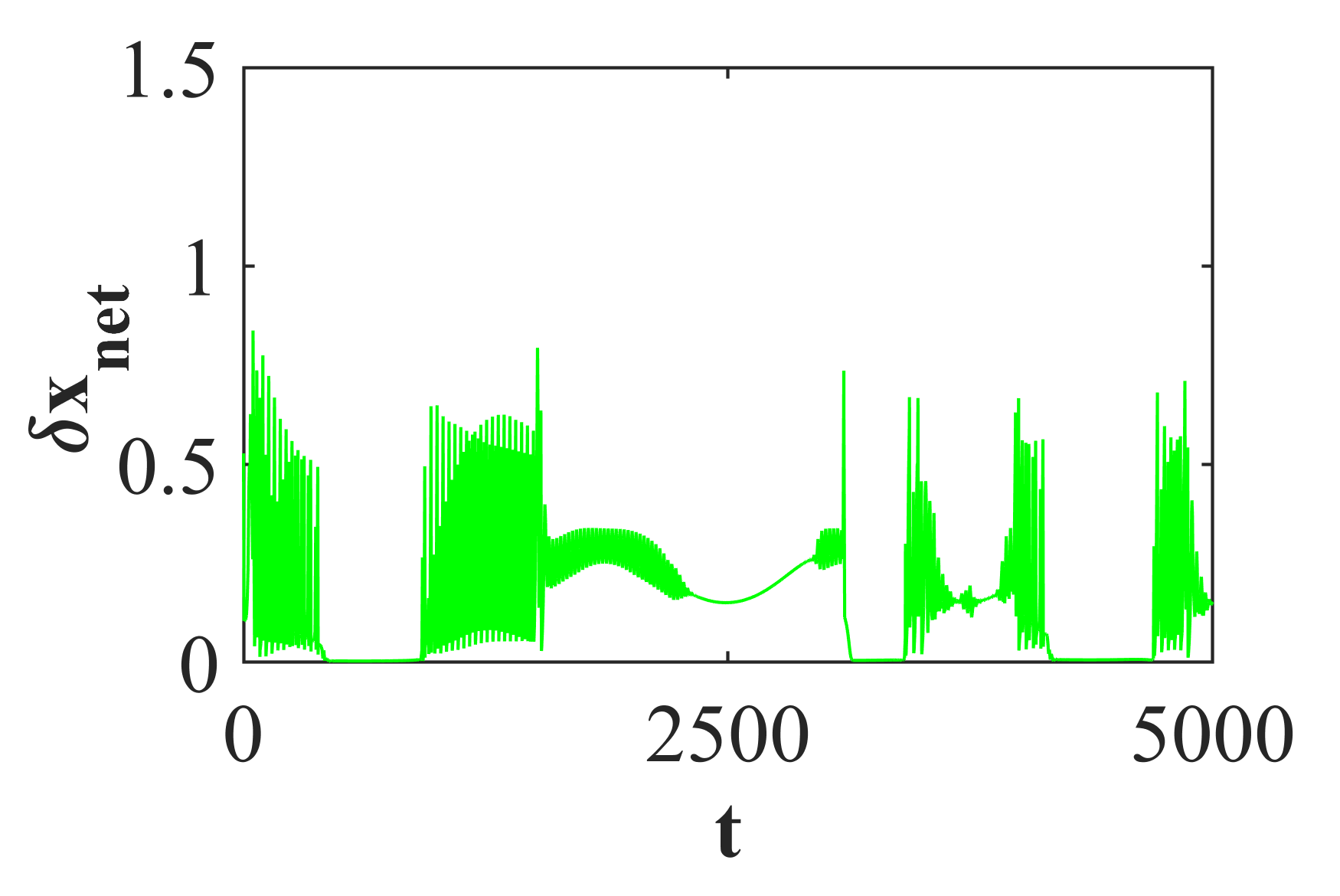} & \includegraphics[width=0.235\textwidth]{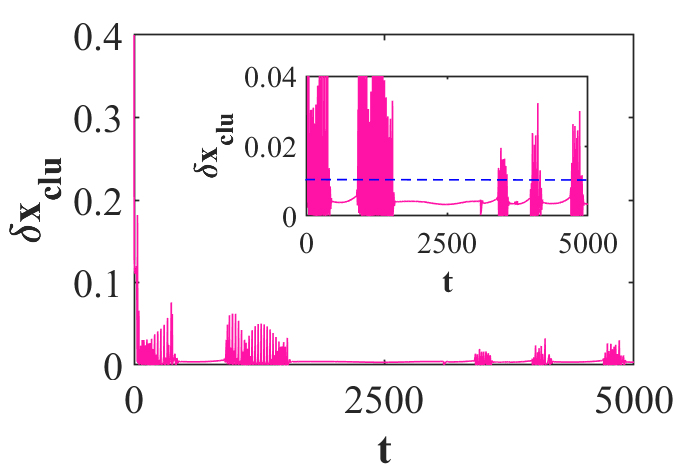} &     \includegraphics[width=0.235\textwidth]{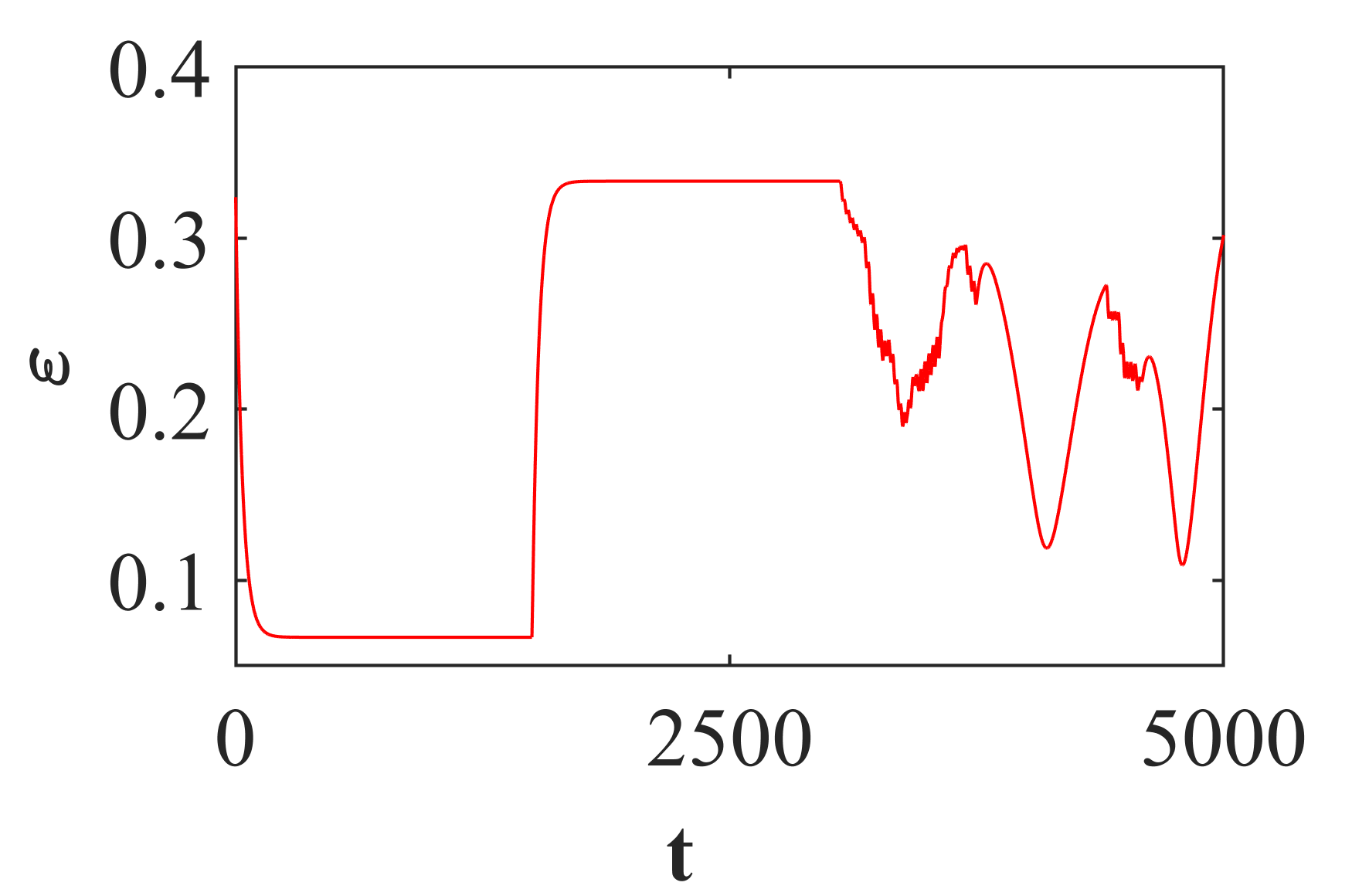}
    \end{tabular}
    \caption{Robustness of breathing cluster synchronization under slower astrocytic feedback delays. The simulations reproduce the dynamics of Fig.~\ref{fig:1}, with only the delay parameter \(\tau\) modified. We reported here the time evolution of the membrane potentials $x(t)$,the  network  synchronization error $\delta x_{net}$, the cluster  synchronization error $\delta x_{clu}(t)$, and the synaptic coupling strength $\epsilon(t)$, for three different feedback delays: $\tau = 500$  (top row), $\tau = 1000$  (middle row), and $\tau = 1500$  (bottom row). The breathing dynamics, characterized by rhythmic alternations between coherent and incoherent states, are preserved, demonstrating robustness beyond the default value \(\tau=200\) used in Fig.~\ref{fig:1}. The blue dashed line in the $\delta x_{\text{clu}}(t)$ panels indicates the desynchronization threshold $\delta_{\text{th}} = 0.01$. Fixed parameters: $\alpha = 0.03$, $\beta = 0.008$, $\gamma = 0.002$, $D = 0.01$.}
    \label{fig:supp1}
\end{figure*}

To assess the robustness of this breathing dynamics with respect to the astrocytic delay, we repeated the simulations of Fig.~\ref{fig:1} by varying only the delay parameter \(\tau\), while keeping all other parameters unchanged. For this purpose, the cases \(\tau=500\), \(1000\), and \(1500\) were considered to account for slower astrocytic calcium dynamics. As shown in Fig.~\ref{fig:supp1}, the characteristic alternation between synchronized and desynchronized cluster states persists across all considered delays, with $\delta x_{\rm clu}(t)$ continuing to alternate between near-zero and elevated values. Increasing $\tau$ mainly induces a phase shift of the adaptive synaptic strength $\varepsilon(t)$ relative to the cluster error, accompanied by only mild changes in the breathing period and switching regularity. These results indicate that the emergence of the breathing pattern is robust to the astrocytic delay and is primarily governed by the adaptive feedback mechanism rather than by its precise timing.
While the present results demonstrate robustness up to $\tau=1500$, astrocytic Ca$^{2+}$ dynamics may occur over substantially longer
timescales. Exploring larger values of $\tau$ will therefore be an important direction for future work, particularly to determine whether slower astrocytic feedback modifies the period, amplitude, or stability of the breathing dynamics.

\section{Representative time series of distinct dynamical regimes}\label{ap2}

\begin{figure*}[htp!]
    \centering
    \begin{tabular}{cccc}
       (a) & (b) & (c) & (d)\\     \includegraphics[width=0.235\textwidth]{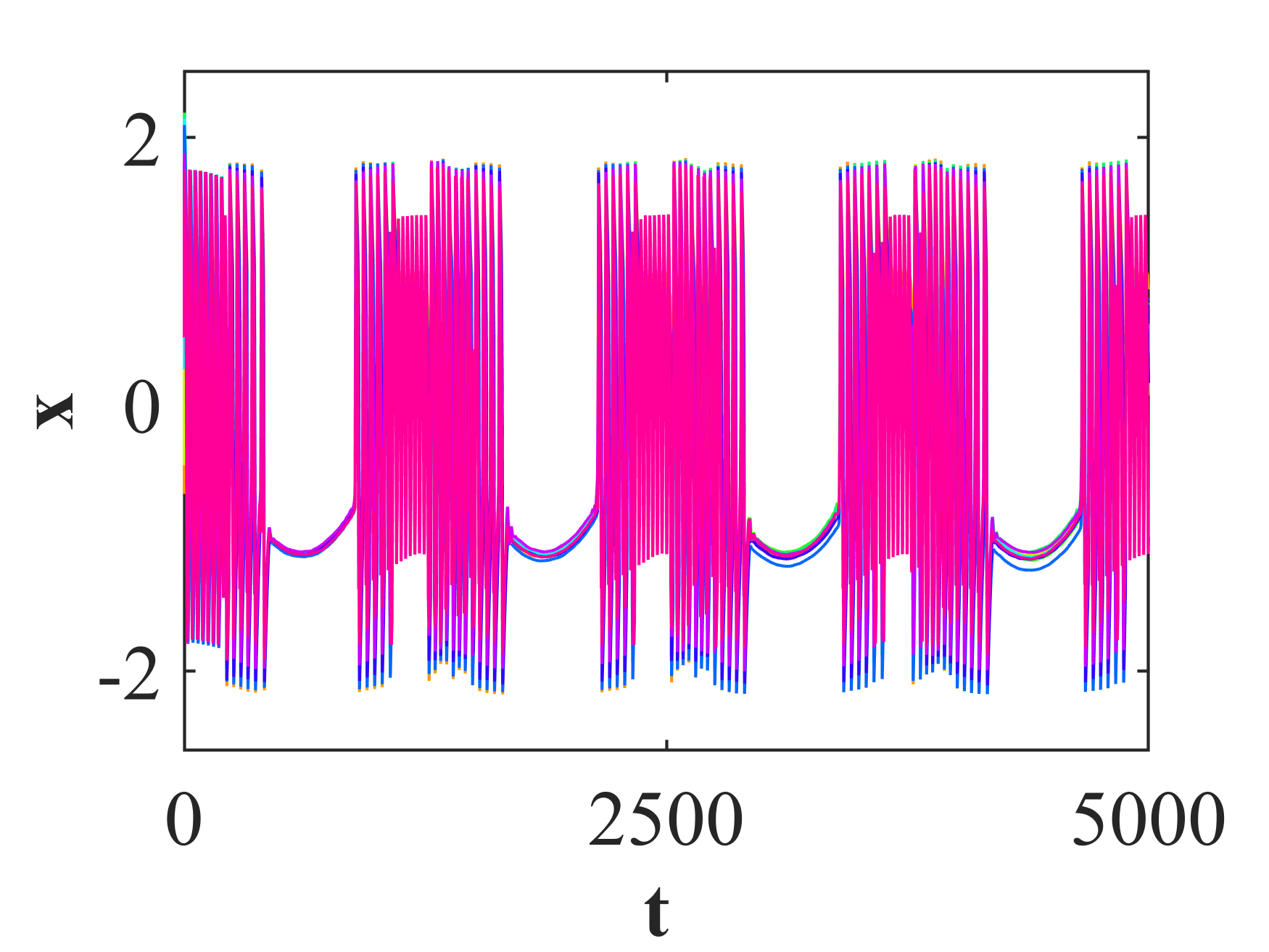} & \includegraphics[width=0.235\textwidth]{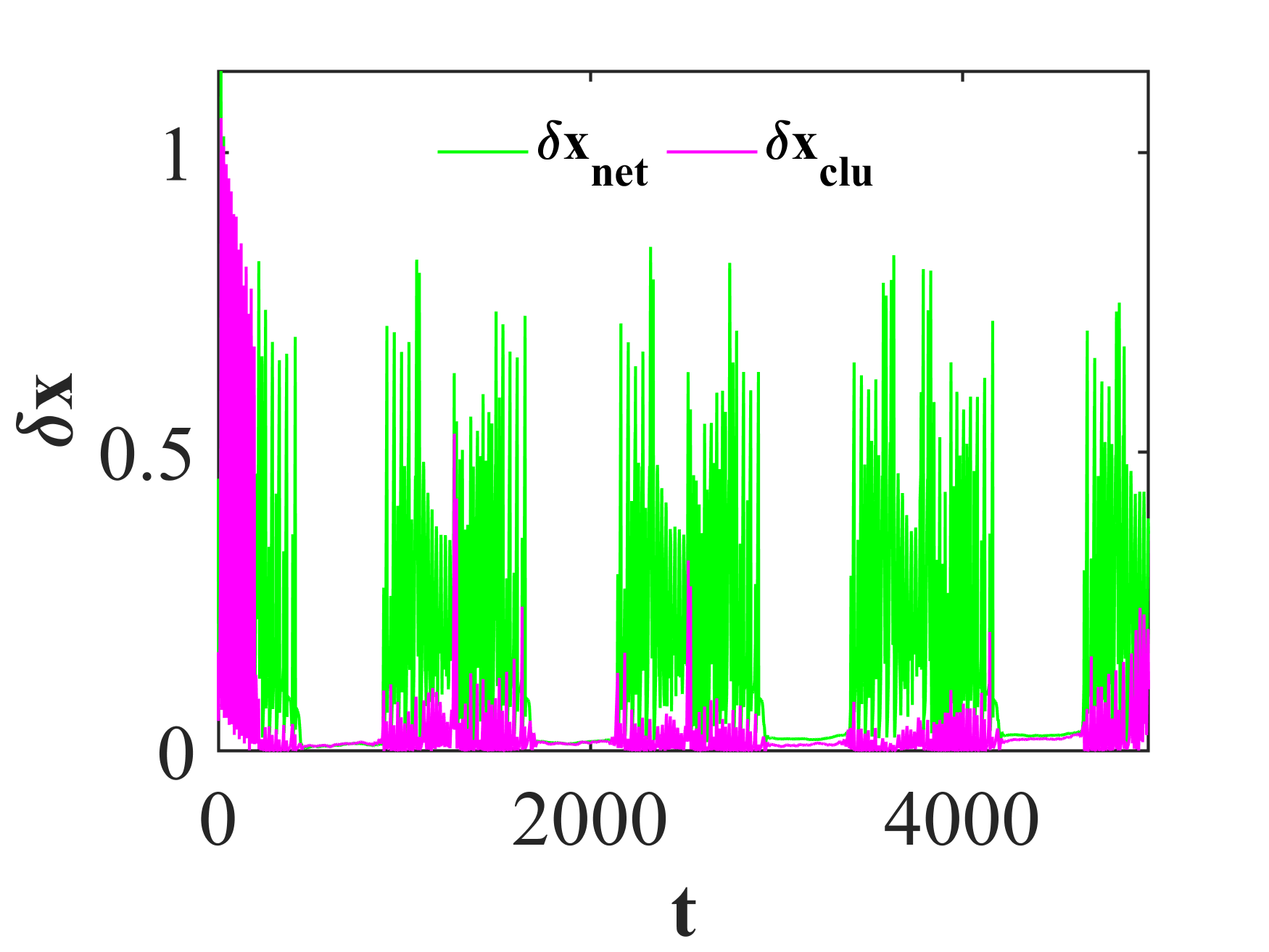}& \includegraphics[width=0.235\textwidth]{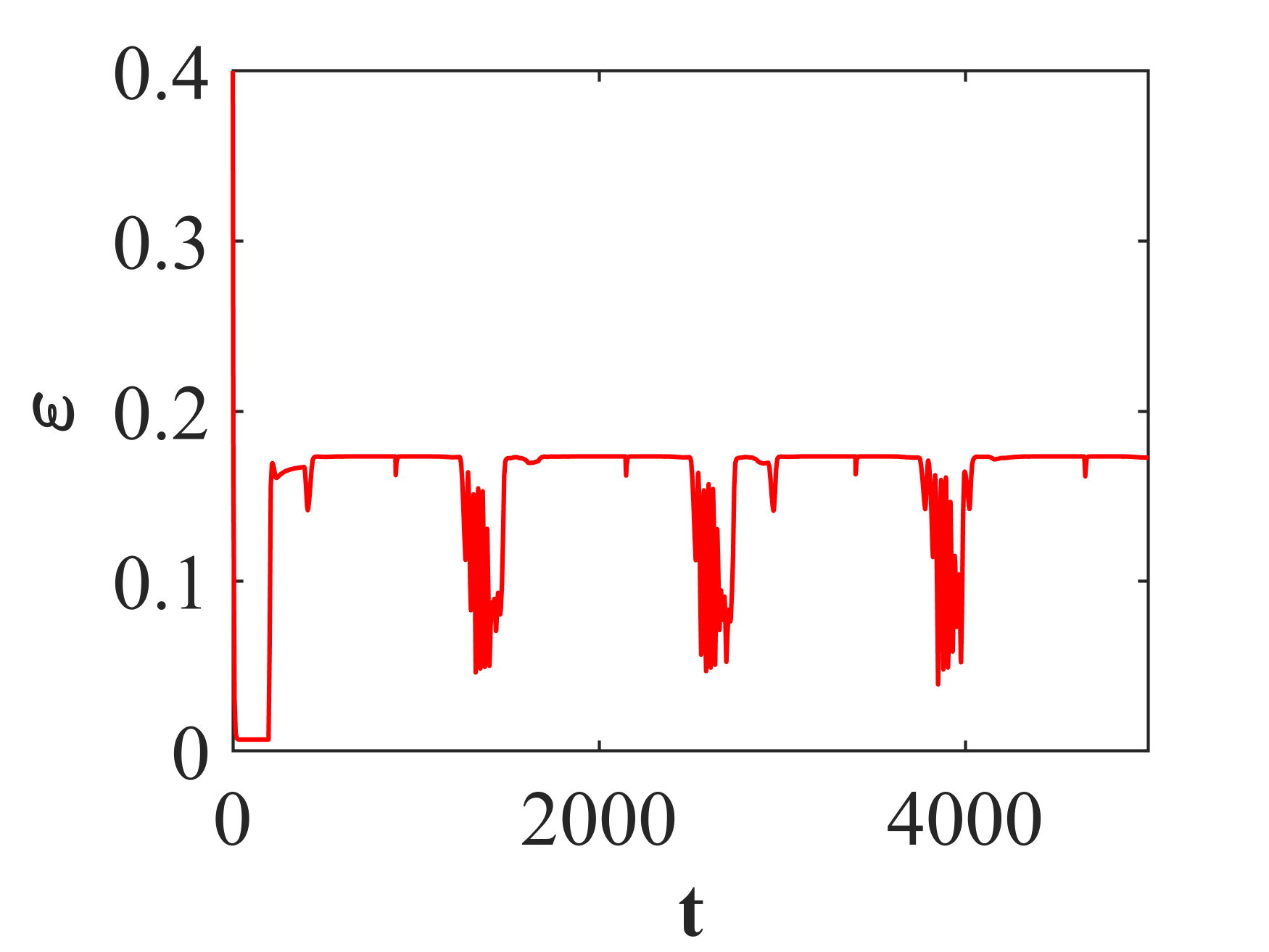} &        \includegraphics[width=0.235\textwidth]{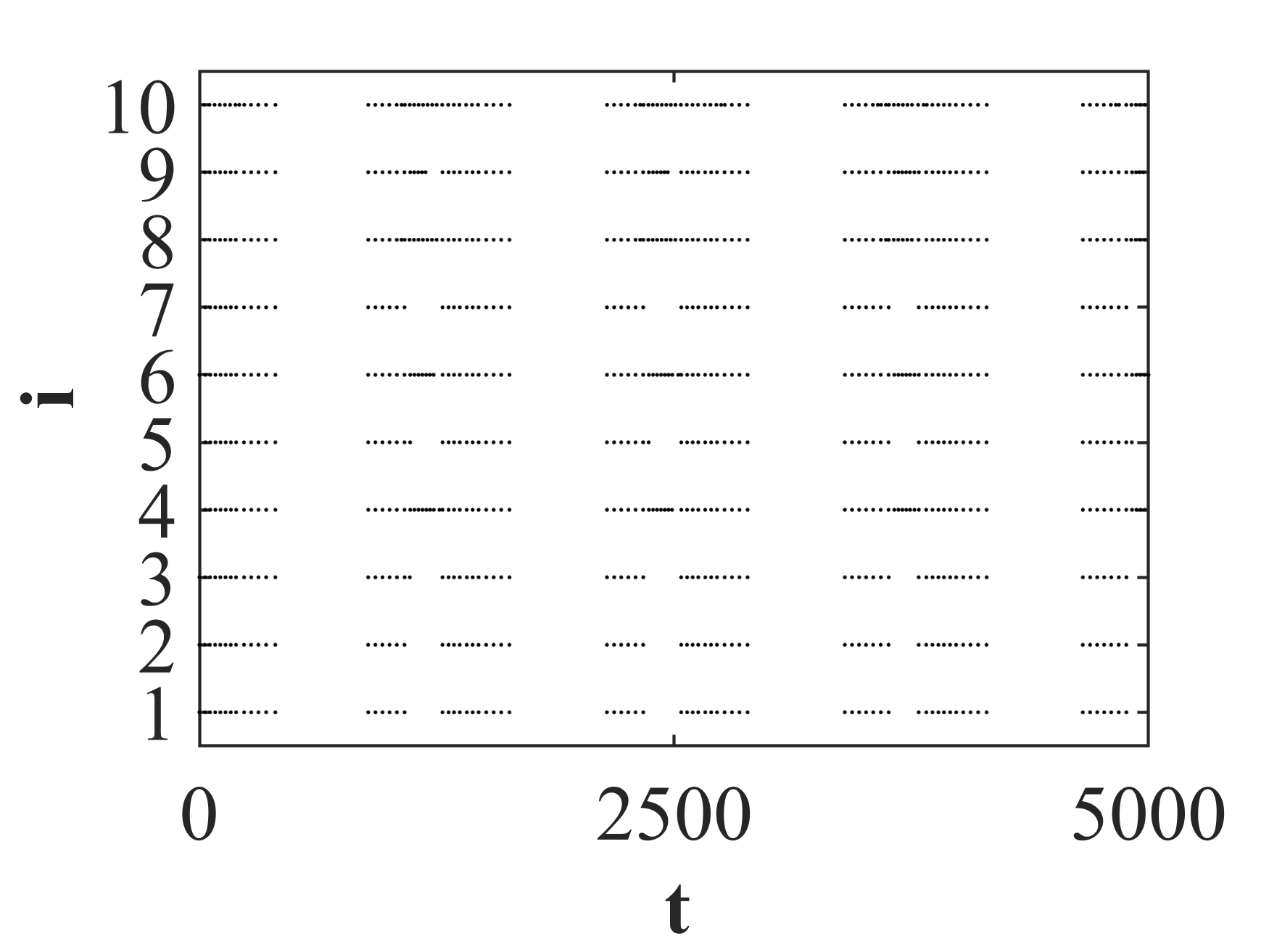}\\
       (e) & (f) & (g) & (h)\\       \includegraphics[width=0.235\textwidth]{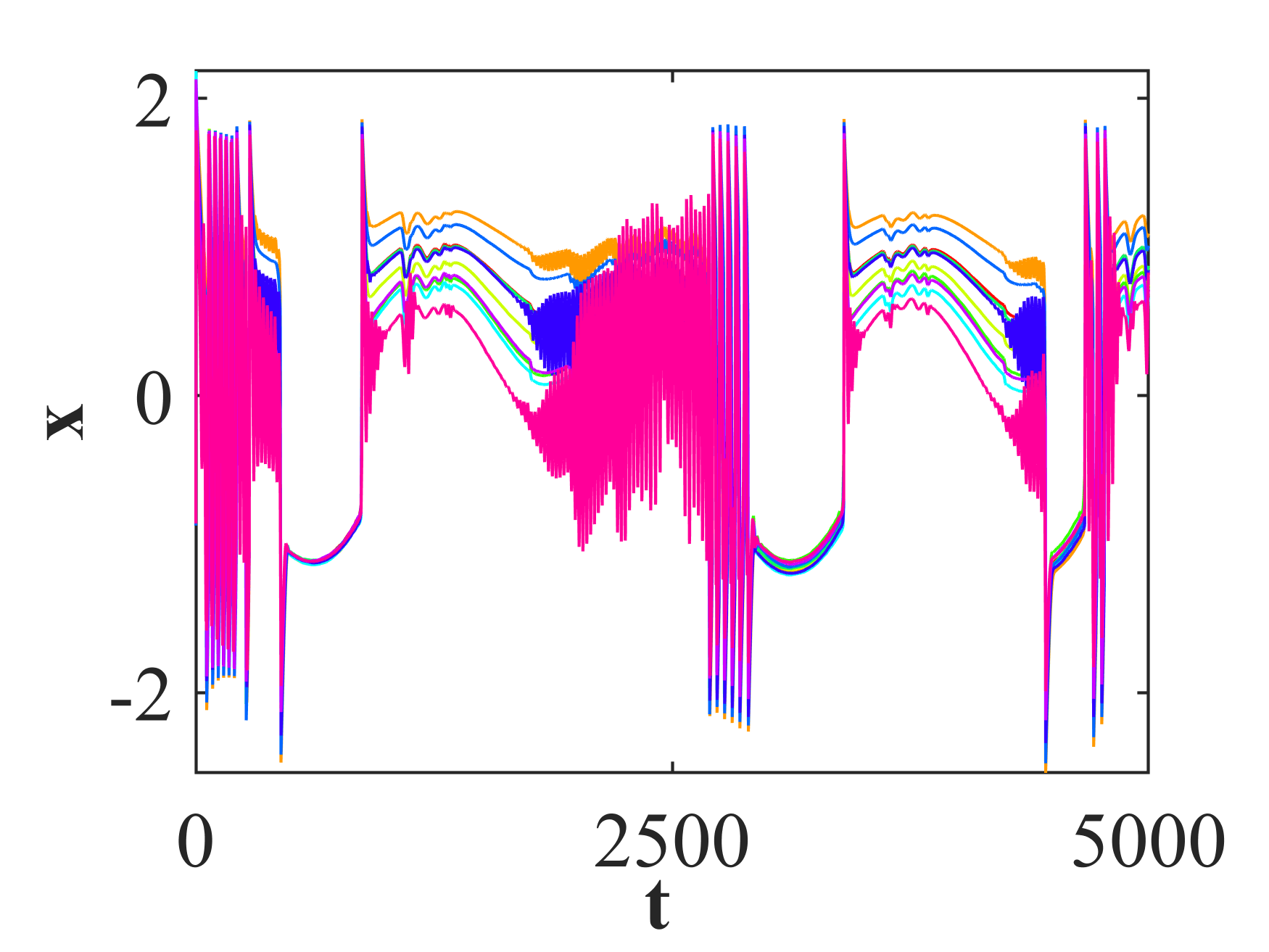}& \includegraphics[width=0.235\textwidth]{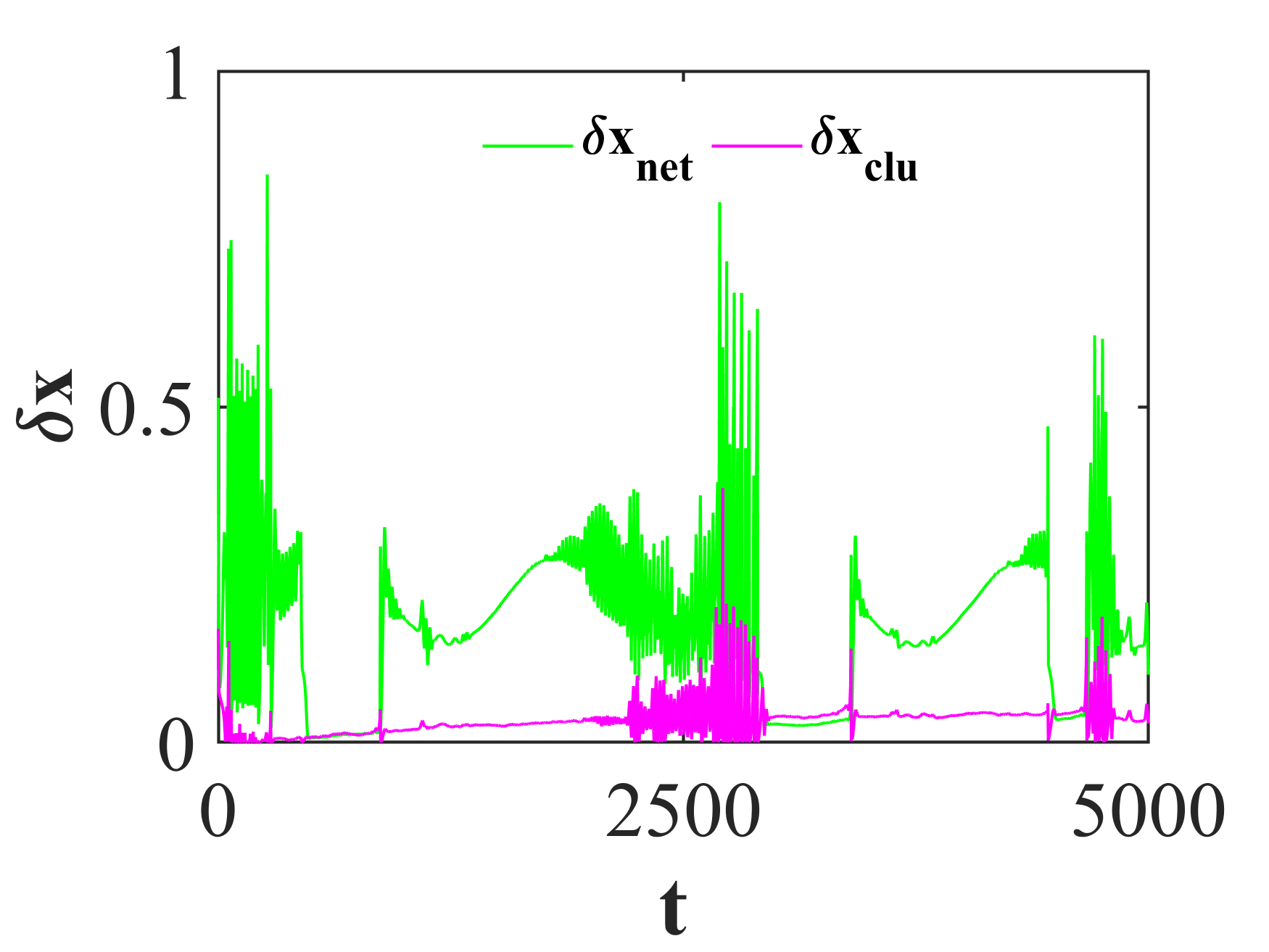} & \includegraphics[width=0.235\textwidth]{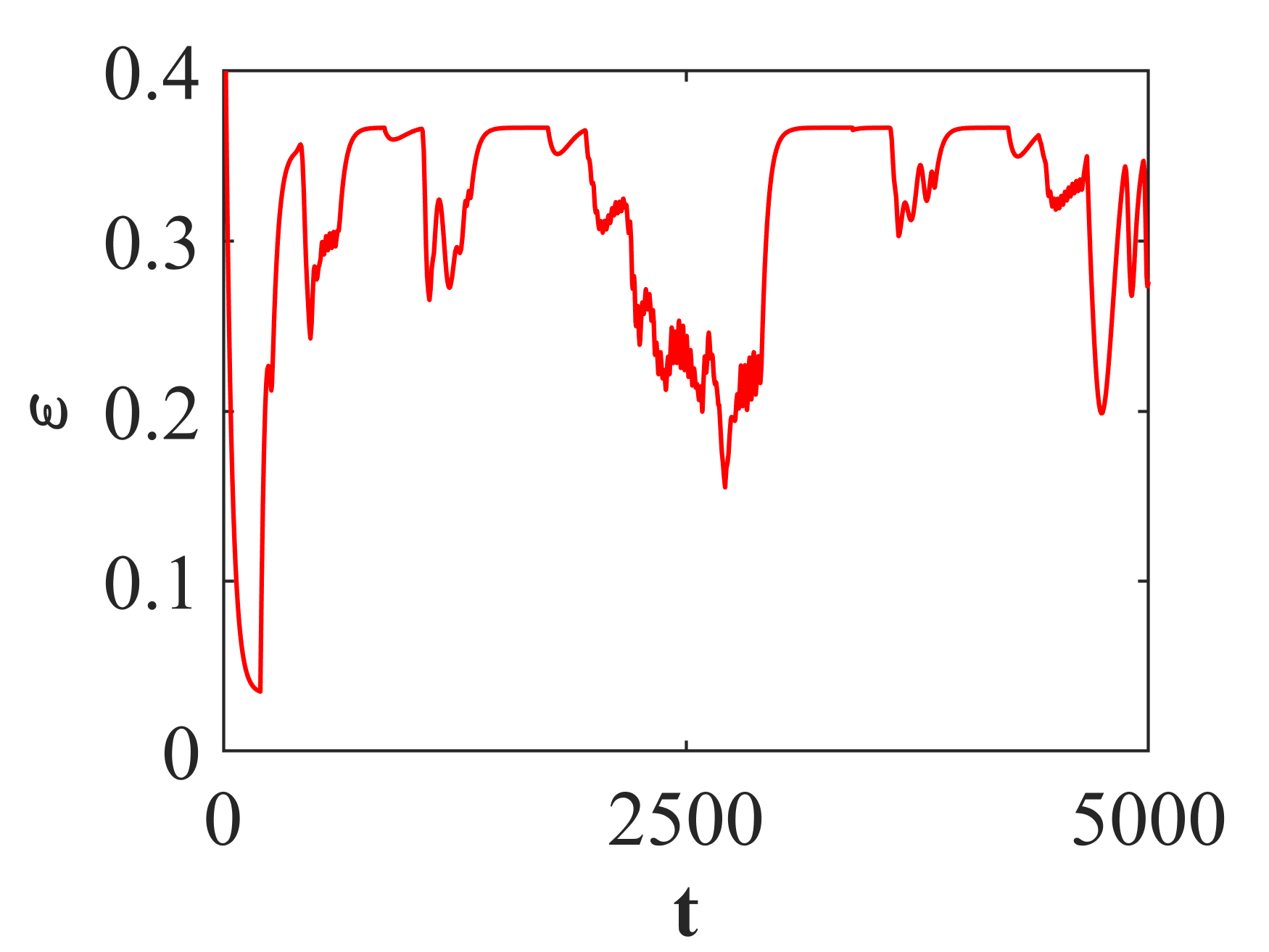} &     \includegraphics[width=0.235\textwidth]{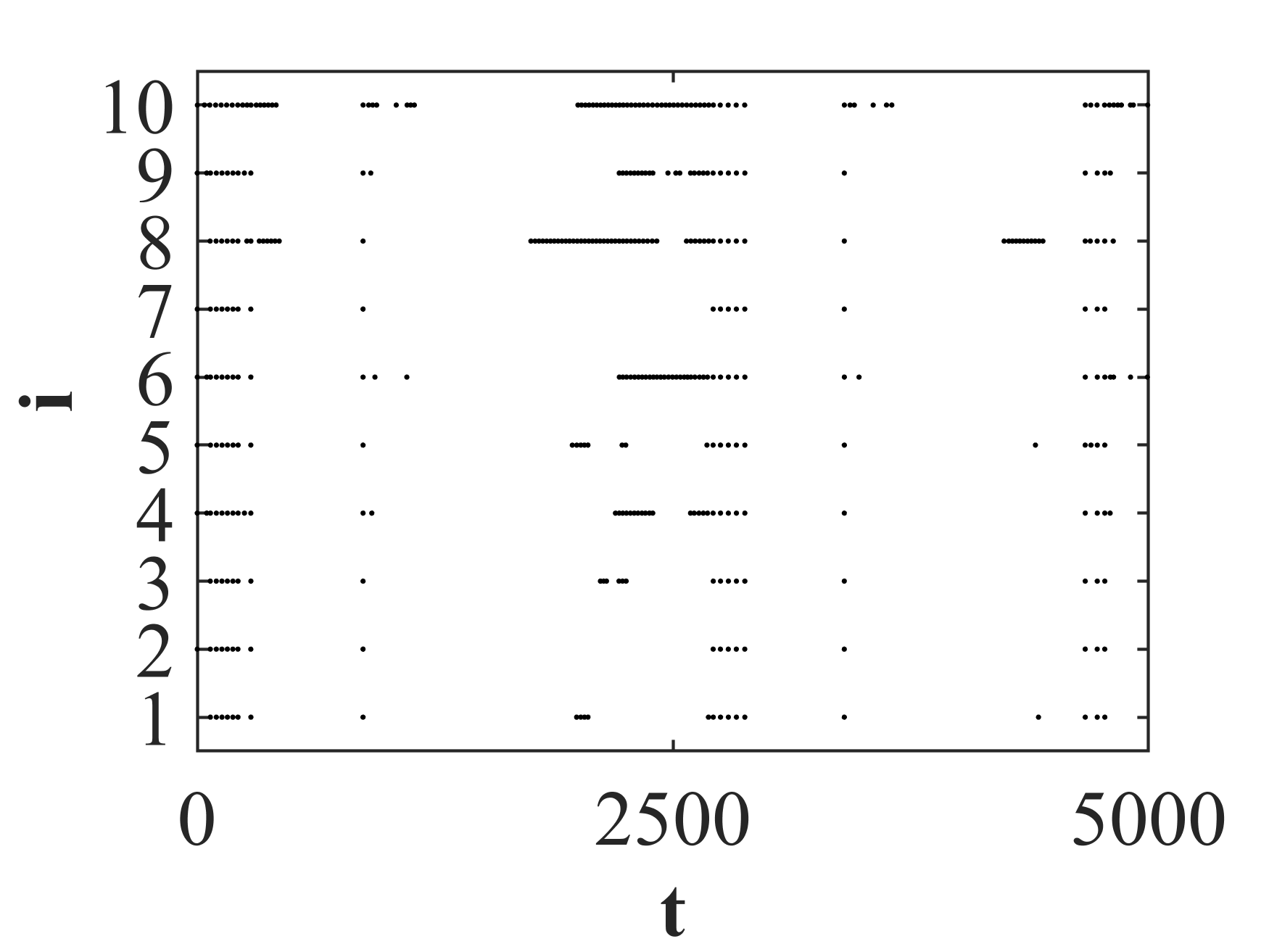}\\
     (i) & (j) & (k) & (l)\\
     \includegraphics[width=0.235\textwidth]{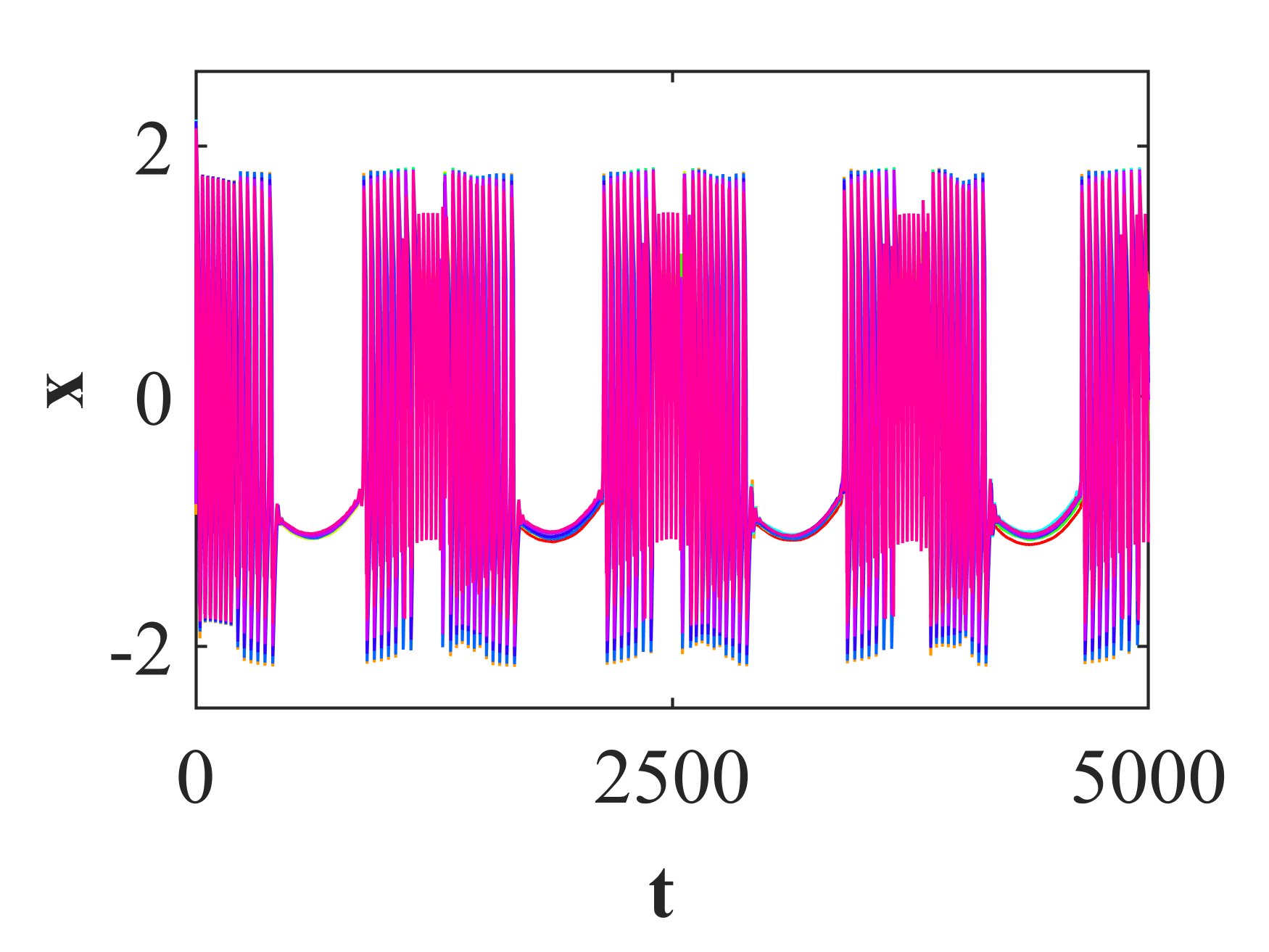}& \includegraphics[width=0.235\textwidth]{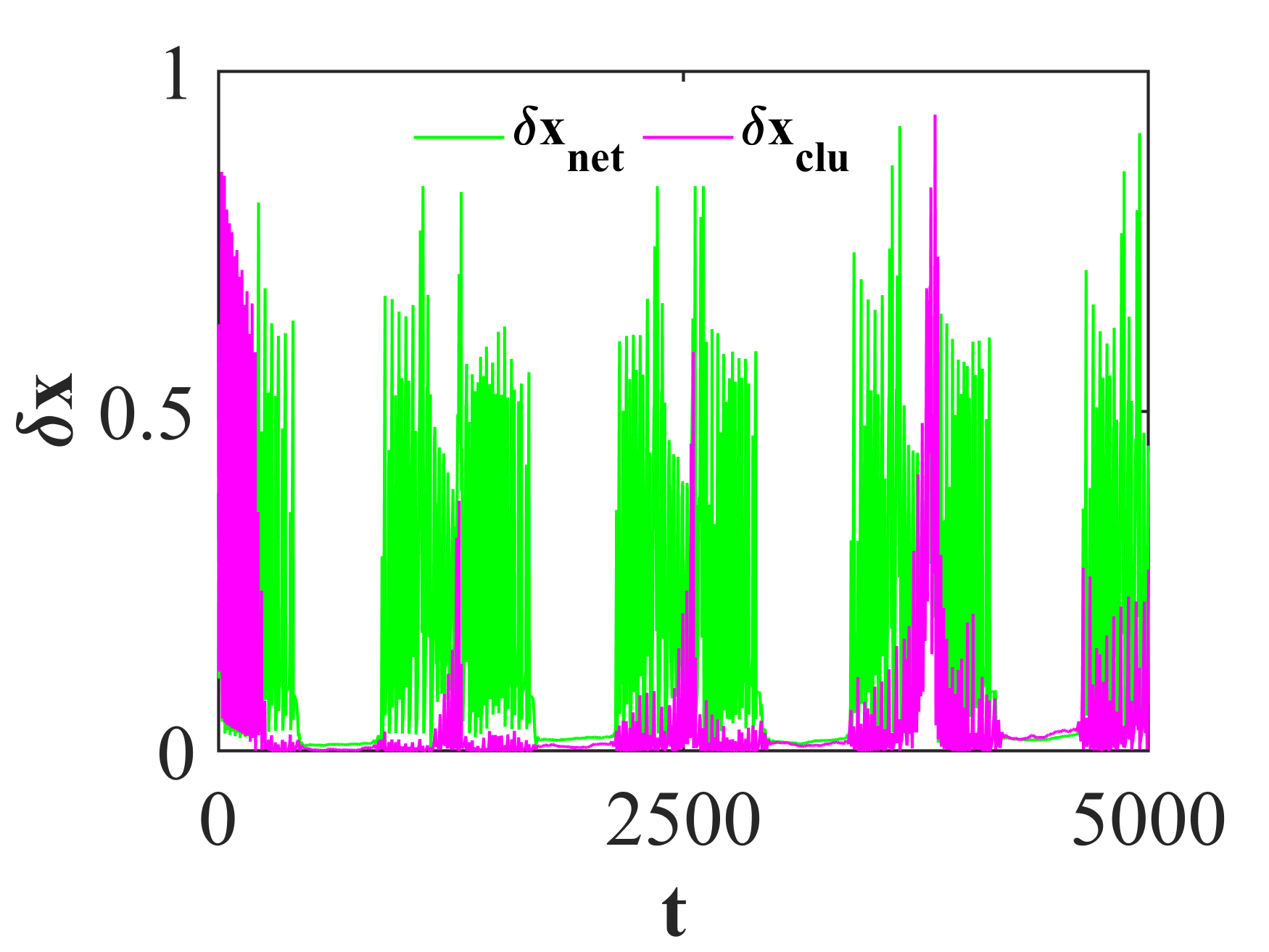} & \includegraphics[width=0.235\textwidth]{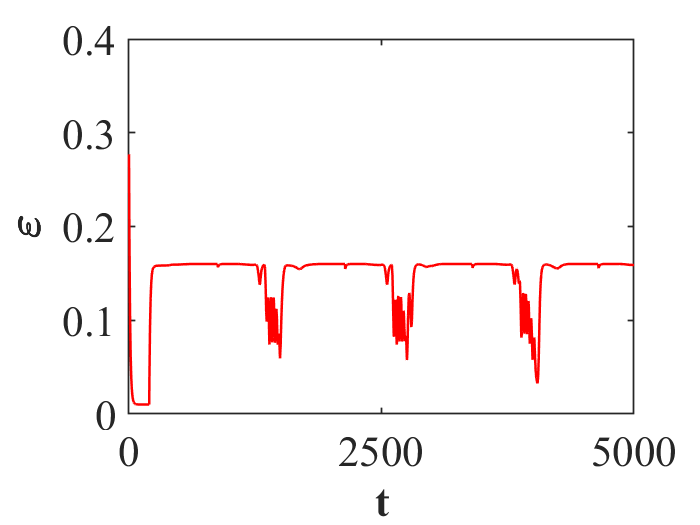} &     \includegraphics[width=0.235\textwidth]{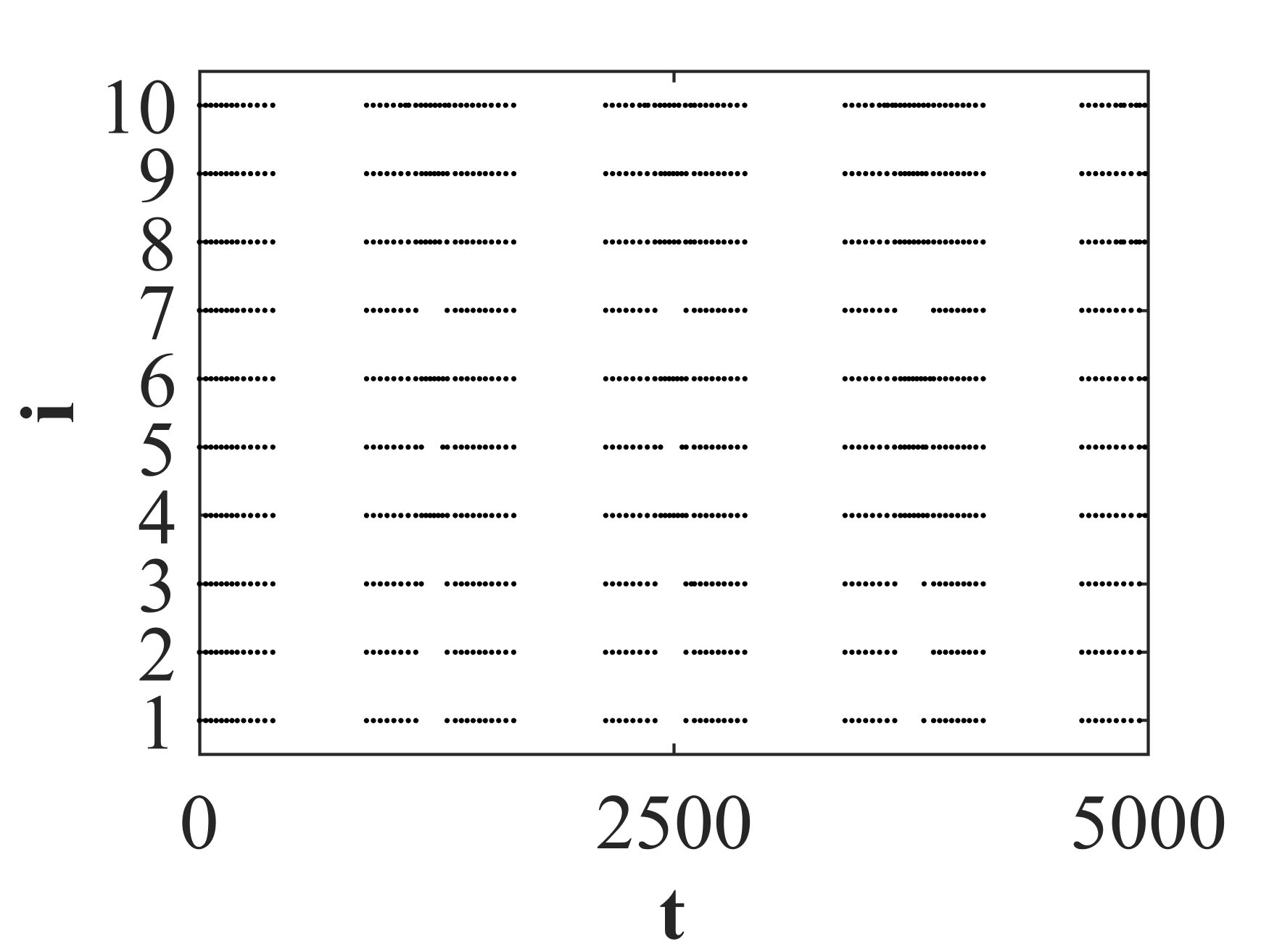}\\
     (m) & (n) & (o) & (p)\\     \includegraphics[width=0.235\textwidth]{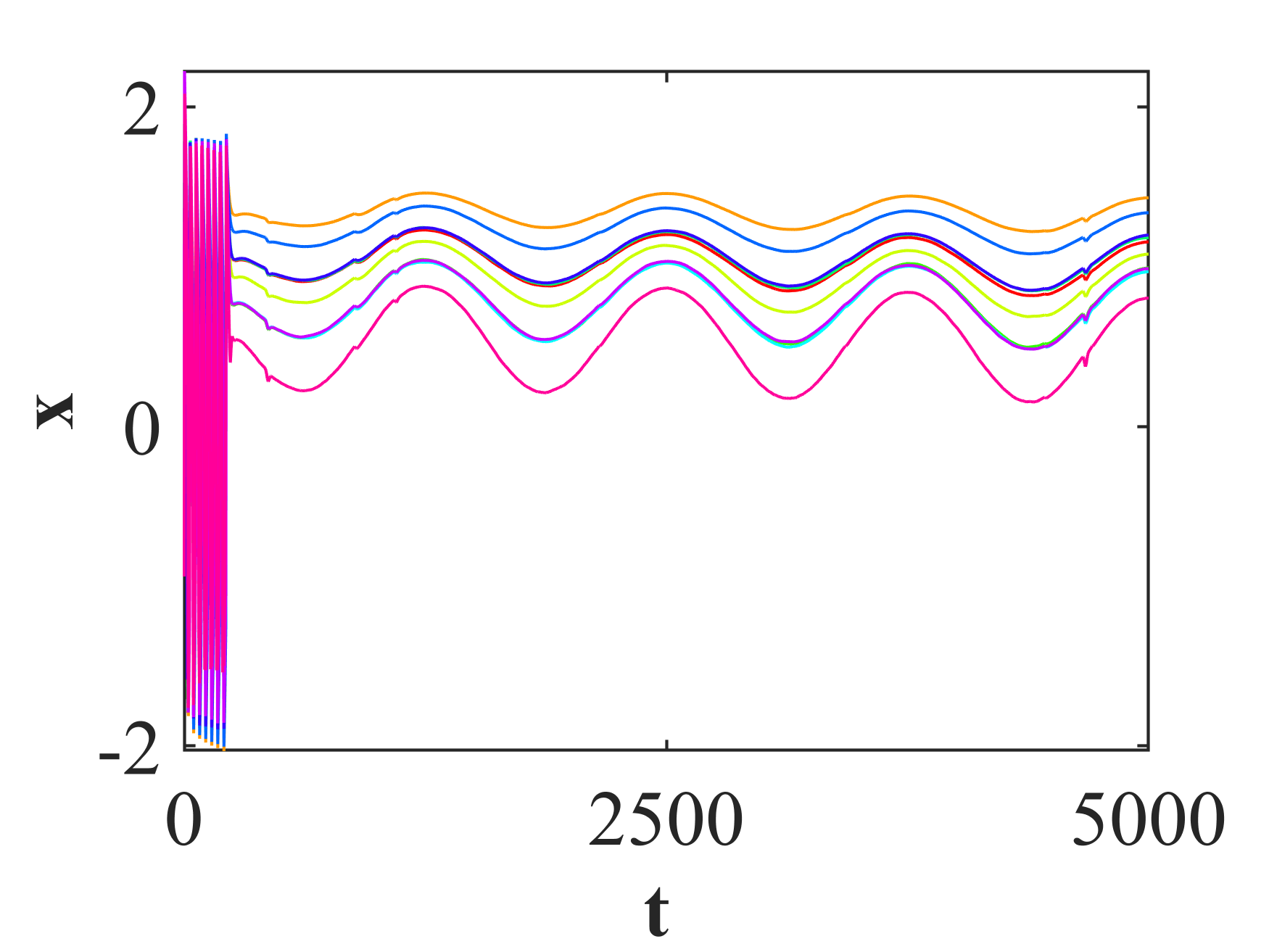} & \includegraphics[width=0.235\textwidth]{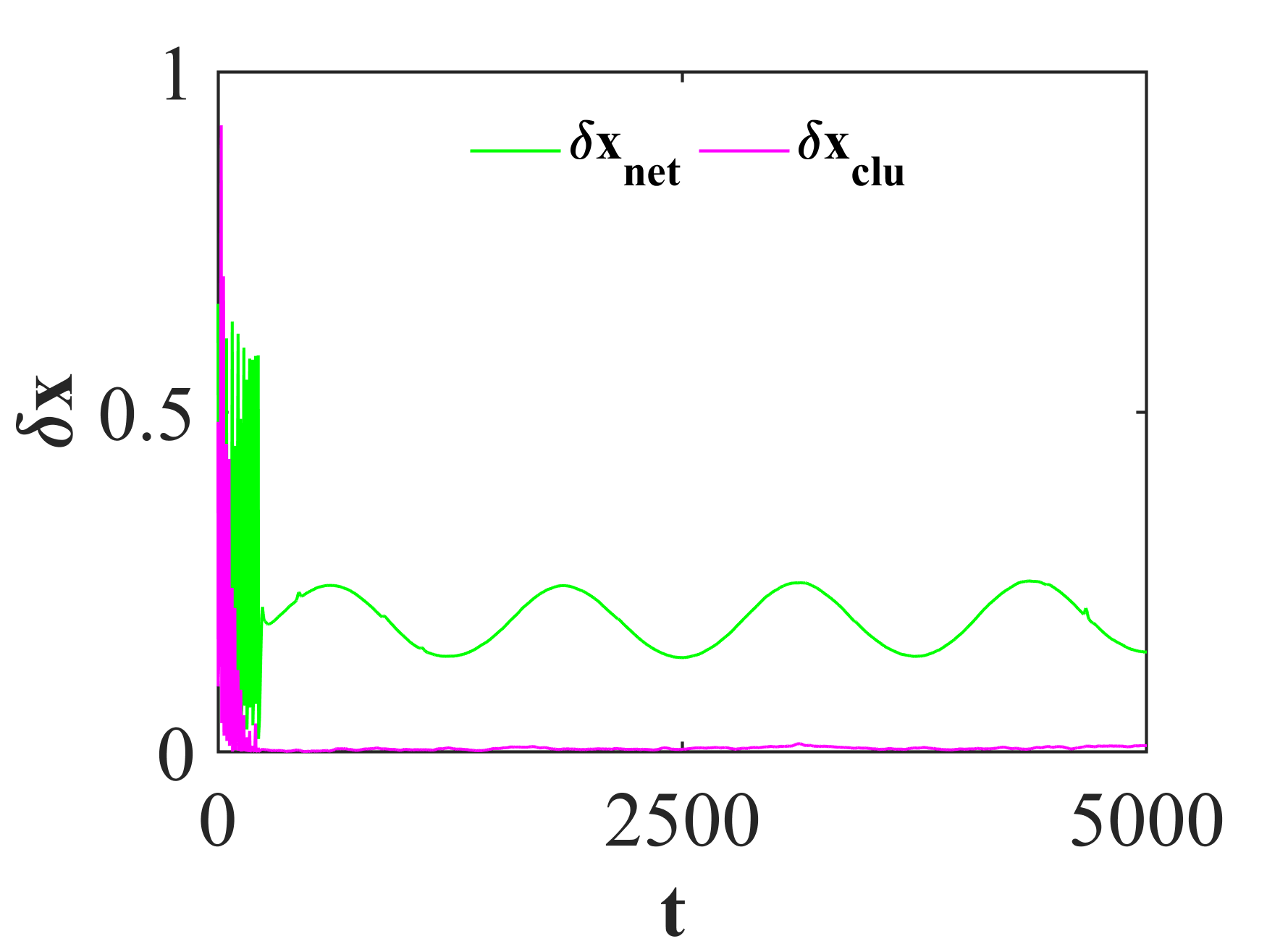}& \includegraphics[width=0.235\textwidth]{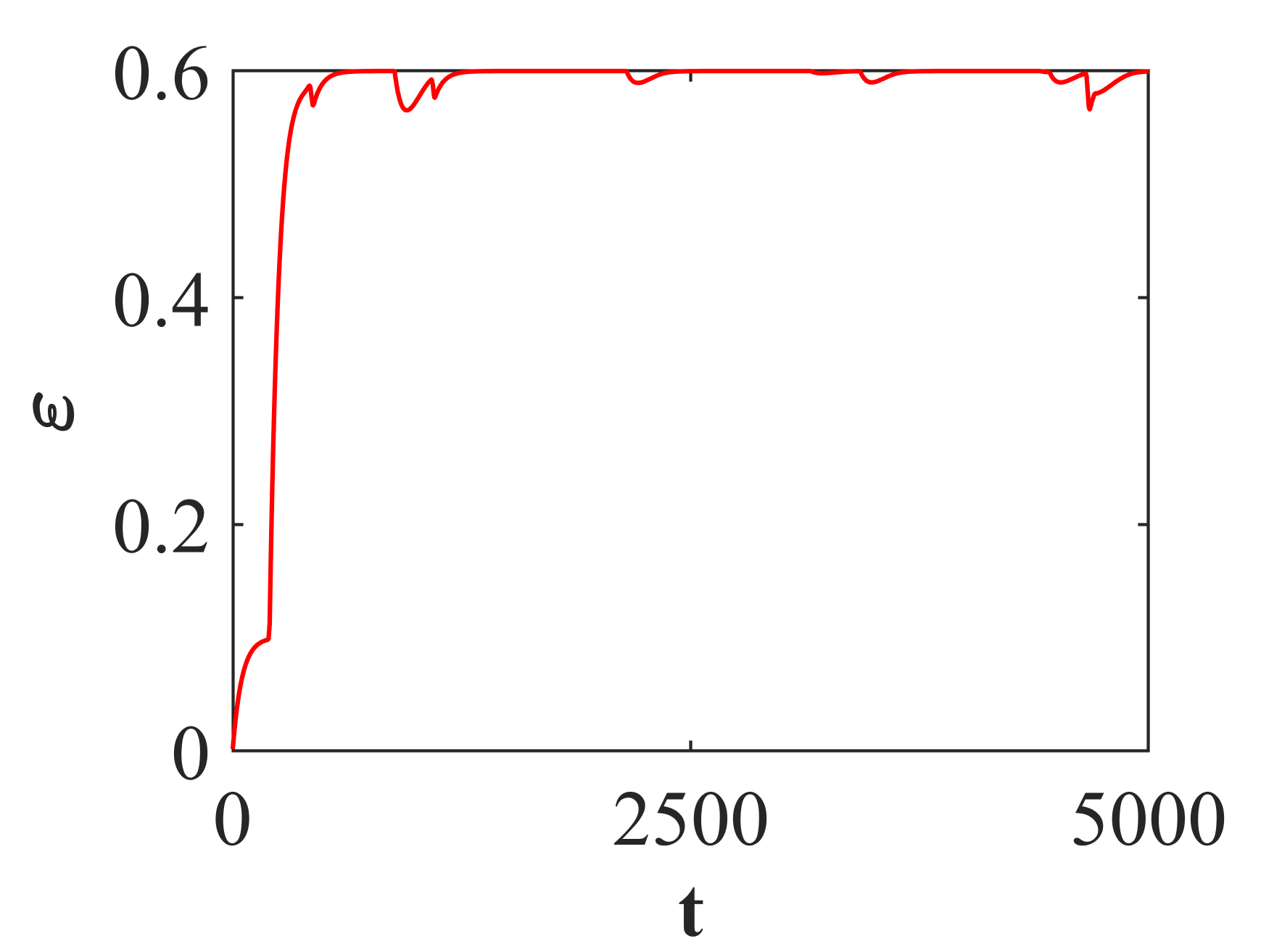} &        \includegraphics[width=0.235\textwidth]{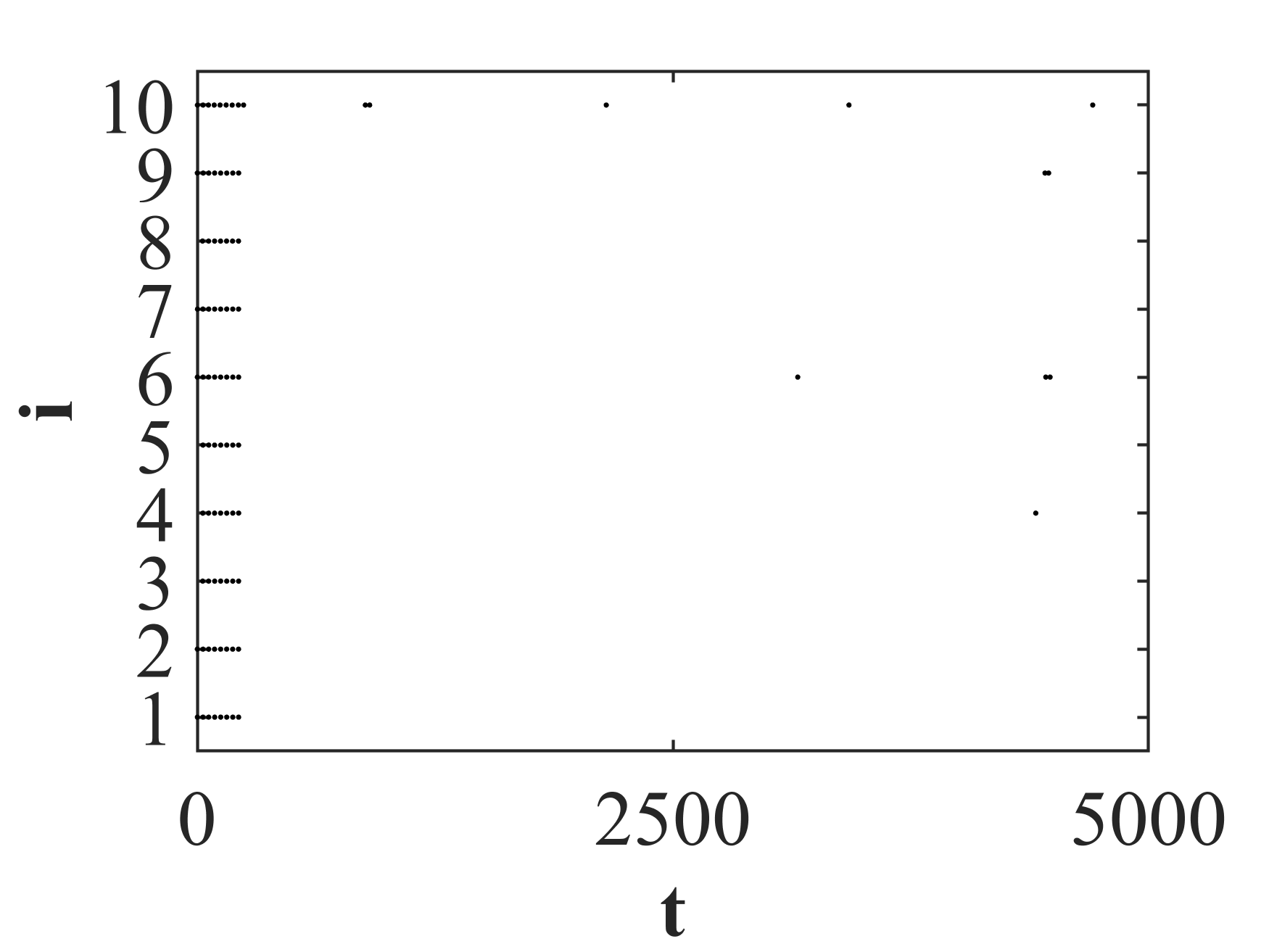}
    \end{tabular}
    \caption{Representative time series illustrating distinct dynamical regimes of the neuron--astrocyte network under high-noise conditions ($D=1$), using parameter sets selected from the high-noise regions of the phase diagrams in Fig.~5. 
The four columns show, respectively, the membrane potentials $x_i(t)$, the global synchronization error $\delta x_{\mathrm{net}}(t)$, the cluster synchronization error $\delta x_{\mathrm{clu}}(t)$ for the symmetric cluster $V_c=\{4,6,9\}$, and the adaptive synaptic coupling strength $\epsilon(t)$. 
The cluster synchronization error $\delta x_{\mathrm{clu}}(t)$ provides the primary diagnostic for distinguishing the dynamical regimes, while the desynchronization fraction $P$ quantifies the fraction of the simulation time for which $\delta x_{\mathrm{clu}}(t)$ exceeds the threshold $\delta_{\mathrm{th}}=0.01$.
Panels (a)--(d) correspond to a moderately desynchronized regime, with $P=0.60$, for $\alpha=0.30$, $\beta=0.05$, and $\gamma=0.002$. In this regime, $\delta x_{\mathrm{clu}}(t)$ remains above the desynchronization threshold for a substantial fraction of the simulation, indicating a predominantly desynchronized cluster with intermittent returns toward coherence. Panels (e)--(h) correspond to a strongly desynchronized regime, with $P=0.85$, for $\alpha=0.30$, $\beta=0.08$, and $\gamma=0.002$. Here, $\delta x_{\mathrm{clu}}(t)$ remains predominantly above the threshold, indicating persistent loss of cluster coherence.
Panels (i)--(l) correspond to a breathing regime, with $P=0.49$, for $\alpha=0.10$, $\beta=0.15$, and $\gamma=0.001$. In this case, $\delta x_{\mathrm{clu}}(t)$ exhibits recurrent alternations between low values below the threshold and elevated values above it, providing a direct signature of the breathing dynamics.
Panels (m)--(p) illustrate the rigidly synchronized regime, with $P=0.04$, for $\alpha=0.02$, $\beta=0.01$, and $\gamma=0.002$, characterized by a persistently low $\delta x_{\mathrm{clu}}(t)$ and the absence of recurrent transitions between synchronized and desynchronized epochs. The corresponding parameter values are given in the text. 
Thus, $P$ is used as a complementary quantitative measure of the temporal prevalence of desynchronization, whereas the temporal evolution of $\delta x_{\mathrm{clu}}(t)$ provides the qualitative distinction between rigid synchronization, breathing synchronization, and persistent desynchronization. 
Unless otherwise stated, the remaining parameters are fixed at $\tau=200$, $T=5$, $a=0.7$, $b=0.4$, $c=0.1$, $\xi=0.175$, $I=0.5$, $k=0.001$, $\omega=0.005$, and $r=0.001$.
} \label{fig:appendix_regimes}
\end{figure*}

To distinguish the dynamical regimes that may exhibit similar values of the desynchronization fraction $P$, we analyse the temporal evolution of the cluster synchronization error $\delta x_{\mathrm{clu}}(t)$, which provides the primary signature of the underlying dynamics. 
Rigid synchronization is characterized by a persistently low cluster synchronization error, i.e., \(\delta x_{\mathrm{clu}}(t)\), indicating sustained cluster coherence. In contrast, breathing synchronization is identified by recurrent alternations between synchronized and desynchronized epochs. 
Persistent desynchronization corresponds to a regime in which \(\delta x_{\mathrm{clu}}(t)\) remains predominantly above the desynchronization threshold. Although the parameter \(P\) quantifies the fraction of the simulation time during which \(\delta x_{\mathrm{clu}}(t)\) exceeds the threshold, it does not uniquely characterize the temporal organization of the dynamics and is therefore used as a complementary measure. Representative time series of the relevant dynamical variables are shown in Fig.~\ref{fig:appendix_regimes} for high-noise conditions ($D=1$), highlighting the distinct temporal signatures associated with rigid synchronization, breathing synchronization, and persistent desynchronization.   


\section*{References}
\bibliography{References}  



\end{document}